\documentclass[letterpaper,titlepage,11pt]{article}
\usepackage{hyperref}

\usepackage{amssymb,amsmath,amsfonts}

\def\clock{{\count0=\time
           \divide\count0 60
           \ifnum\count0<10 0\fi\the\count0
           \multiply\count0 -60 \advance\count0 \time
           :\ifnum\count0<10 0\fi \the\count0
         }}
\newcommand{\timestamp}{{\small\vbox{\hbox{\tt\jobname.tex}
\hbox{\the\day/\the\month/\the\year, \clock}}}}

\def\BB{{\cal B}}

\def\TT{{\cal T}}

\def\WW{{\cal W}}

\newcommand{\beq}{\begin{equation}}
\newcommand{\eeq}{\end{equation}}
\newcommand{\ben}{\begin{displaymath}}
\newcommand{\een}{\end{displaymath}}
\newcommand{\beqa}{\begin{eqnarray}}
\newcommand{\eeqa}{\end{eqnarray}}
\newcommand{\bea}{\begin{eqnarray}}
\newcommand{\eea}{\end{eqnarray}}
\newcommand{\bean}{\begin{eqnarray*}}
\newcommand{\eean}{\end{eqnarray*}}
\newcommand{\ba}{\begin{array}}
\newcommand{\ea}{\end{array}}
\newcommand{\bi}{\begin{itemize}}
\newcommand{\ei}{\end{itemize}}
\newcommand{\ie}{{\it i.e.,\,}}
\newcommand{\eg}{{\it e.g.,\,}}

\newcommand{\vep}{\varepsilon}
\newcommand{\Ld}{\pounds}
\newcommand{\oln}{\overline\nabla}

\numberwithin{equation}{section}

\begin{document}

\begin{titlepage}
\begin{flushright}
CPHT-RR095.1009\\
NORDITA-2009-65 \\
\end{flushright}
\vskip 1.8cm
\begin{center}
{\bf\LARGE{Essentials of Blackfold Dynamics}} 
\vskip 1.6cm
{\bf 
Roberto Emparan$^{a,b}$, Troels Harmark$^{c,d}$,
Vasilis Niarchos$^{e}$,}
{\bf
Niels A. Obers$^{c}$
}
\vskip 0.5cm
\medskip
\textit{$^{a}$Instituci\'o Catalana de Recerca i Estudis
Avan\c cats (ICREA)}\\
\smallskip
\textit{$^{b}$Departament de F{\'\i}sica Fonamental, Universitat de
Barcelona, }\\
\textit{Marti i Franqu\`es 1, E-08028 Barcelona, Spain}\\
\smallskip
\textit{$^{c}$The Niels Bohr Institute,}
\textit{Blegdamsvej 17, 2100 Copenhagen \O, Denmark}\\
\smallskip
\textit{$^{d}$NORDITA,
Roslagstullsbacken 23,
SE-106 91 Stockholm,
Sweden}\\
\smallskip
\textit{$^{e}$Centre de Physique Th\'eorique, \'Ecole Polytechnique,
 91128 Palaiseau, France}\\
 \textit{Unit\'e mixte de Recherche 7644, CNRS}

\vskip .2 in
{\tt emparan@ub.edu, harmark@nordita.org, niarchos@cpht.polytechnique.fr,
obers@nbi.dk}
\end{center}
\vskip 0.3in

\baselineskip 16pt
\date{}

\begin{center}
{\bf Abstract}
\end{center}
\vskip 0.2cm \noindent 

We develop and significantly generalize the effective worldvolume theory
for higher-dimensional black holes recently proposed by the authors. The
theory, which regards the black hole as a black brane curved into a
submanifold of a background spacetime ---a {\it blackfold}---, can be
formulated in terms of an effective fluid that lives on a dynamical
worldvolume. Thus the blackfold equations split into intrinsic
(fluid-dynamical) equations, and extrinsic (generalized geodesic
embedding) equations. The intrinsic equations can be easily solved for
equilibrium configurations, thus providing an efficient formalism for
the approximate construction of novel stationary black holes.
Furthermore, it is possible to study time evolution. In particular, the 
long-wavelength component of the Gregory-Laflamme instability of black 
branes is obtained as a sound-mode instability of the effective fluid. We also
discuss action principles, connections to black hole thermodynamics, and
other consequences and possible extensions of the approach. Finally, we
outline how the fluid/AdS-gravity correspondence is related to this formalism.

\end{titlepage} \vfill\eject

\setcounter{equation}{0}

\pagestyle{empty}
\small
\tableofcontents
\normalsize
\pagestyle{plain}
\setcounter{page}{1}

\newpage


\section{Introduction}

In recent work we have identified the origin of the rich variety of
higher-dimensional black holes in the possibility of having horizons
that are much longer along some directions than in others
\cite{Emparan:2009cs}.
In such cases there are at least two different horizon length-scales, and
we have proposed an effective theory that captures the long-distance
physics when these scales are widely separate. Focusing on the simplest,
universal case of neutral, vacuum black holes, the two length scales
are associated with the mass and angular
momentum\footnote{$J=(\sum_i J_i^2)^{1/2}$ aggregates the effect of all
possible angular momenta.}, 
\beq\label{scales}
\ell_M\sim(GM)^\frac{1}{D-3}\,, \qquad \ell_J\sim \frac{J}{M}\,. 
\eeq 
In a four-dimensional black hole, by virtue of the Kerr bound $J\leq
GM^2$ these lengths are always parametrically similar, but
higher-dimensional black holes (including also black rings) with
$\ell_J\gg \ell_M$ are known to exist in ultra-spinning regimes where
the angular momentum for a given mass can be arbitrarily high
\cite{Myers:1986un,Emparan:2003sy,Emparan:2001wn,Emparan:2008eg}. It
appears that essentially all their novel features, compared to their
four-dimensional cousins, arise from the ability to separate these two
lengths. This suggests that higher-dimensional black holes must be
organized according to a hierarchy of scales:

\begin{enumerate}

\item $\ell_J \lesssim\ell_M$: black holes behave
qualitatively similarly to the four-dimensional Kerr black hole.

\item $\ell_J \approx \ell_M$: threshold of new black hole dynamics. 

\item $\ell_J\gg \ell_M$: the separation of scales suggests an
effective description of long-wavelength dynamics.

\end{enumerate}

The first and second regimes fully involve the non-linearities of
General Relativity, but in the first case we have
no hints of qualitatively new properties of black holes compared to
four-dimensional ones. In particular we conjecture that for
$J<J_\mathrm{crit}=\alpha_D M(GM)^\frac{1}{D-3}$ (with
$\alpha_D$ a yet-undetermined numerical constant of order one),
the Myers-Perry black holes are
dynamically stable and unique among solutions with connected regular
horizons\footnote{This extends a previous conjecture in \cite{Emparan:2003sy} about
stability of MP black holes to
involve their uniqueness too. Here $J_\mathrm{crit}$ refers to the minimum $J$ for
which either instability or 
non-uniqueness first appear.}. In contrast, as the two scales begin to
diverge
in
the second
regime, we have good evidence of the onset of new phenomena: horizon
instabilities, inhomogeneous (`pinched') phases, non-spherical horizon
topologies, and absence of uniqueness
\cite{Emparan:2001wn,Emparan:2003sy,Emparan:2007wm,Dias:2009iu}. This
regime seems hard to investigate by means of exact analytical
techniques, but on the other hand the presence of only one scale in the
problem is actually convenient for numerical investigation, since one
does not require high precision over widely different scales. In the
third regime, which is the focus of this paper, the existence of a small
parameter $\ell_M/\ell_J$ allows the introduction of efficient
approximate analytical methods. In this way, this organization in
scales\footnote{The approach to extremality in a black hole introduces a
long length scale {\it transverse} to the horizon and allows to decouple
a different sector of the physics, namely the near-horizon region
\cite{Kunduri:2007vf}.} provides an outline for a program to investigate
higher-dimensional black holes\footnote{See
\cite{Obers:2008pj,Niarchos:2008jc} for brief reviews of
higher-dimensional black holes and
\cite{Emparan:2008eg} for a more extensive one.}.

We base the effective description of black holes in regime 3 on the idea
that the limit $\ell_M/\ell_J\to 0$ of the black hole, keeping its
horizon size finite, results in a flat, infinitely extended black brane,
possibly boosted along some of its worldvolume directions. This has been
observed in all exact solutions known so far: the horizon of Myers-Perry
black holes pancakes along the plane of rotation and in the limit
becomes a black brane \cite{Emparan:2003sy}, and five-dimensional black
rings become thin and locally approach the geometry of boosted black
strings \cite{Elvang:2003mj,Emparan:2006mm}. This allows us to identify
the variables for the effective description. Essentially, the black hole
will be regarded as a {\it black} brane whose worldvolume spans a curved
submani{\it fold} of a background spacetime --- what we refer to as a
{\it blackfold}. The simplest example is the characterization of a
(thin) black ring as a circular boosted black string, which was worked
out in detail in \cite{Emparan:2007wm} and expanded upon in
\cite{Caldarelli:2008pz,Camps:2008hb}. But, as shown in
\cite{Emparan:2009cs}, ultraspinning Myers-Perry black holes, as well as
a number of other new black holes, can also be appropriately captured
with this approach. 

Thus, the effective theory of black holes when $\ell_M/\ell_J\ll 1$ is a
theory that describes how to bend the worldvolume of a black brane in a
background spacetime. In this regard we treat black branes in a manner similar
to other familiar extended objects such as
cosmic strings or D-branes. The main novelty is that black branes possess
black hole horizons, so when their worldvolume is spatially compact we
obtain a black hole with finite horizon area. 

To leading order in the expansion in $\ell_M/\ell_J$ the backreaction of
the blackfold on the geometry is neglected --- it is a `test' blackfold.
Corrections to the geometry are found by first computing the linearized
gravitational backreaction on the background, which is of order
$(\ell_M/\ell_J)^{D-3}$, then this correction induces a perturbation of the
near-horizon geometry, and so forth. This can be systematically pursued
in the form of a matched asymptotic expansion. For higher-dimensional
black rings these first corrections have been computed
\cite{Emparan:2007wm}. In this paper, however, we remain at the `test
blackfold' level of approximation. This is enough to reveal new kinds of
black holes, compute their physical properties, and also study
time-dependent situations and stability.

Ref.~\cite{Emparan:2009cs} gave a basic outline of the theory of blackfolds. Here we
develop its conceptual basis and considerably improve and generalize the
presentation. Its application to specific new classes of
higher-dimensional black holes will be discussed in a forthcoming
publication. 

We draw heavily from the beautiful theory of classical brane dynamics
developed by Carter in \cite{Carter:2000wv}\footnote{Our title deliberately
highlights this.}. In this respect, we may say that a main part of our
contribution is to apply this theory to \textit{black} branes, and to
interpret the results in the context of higher-dimensional black hole
physics according to the general considerations above. But we
also emphasize that the general classical brane dynamics, regarded as a
long-wavelength effective theory, must take the form of the dynamics of
a fluid that lives on a dynamical worldvolume. Black branes correspond
to a specific type of fluid, with a certain equation of state and with
specific values for transport coefficients. At the leading order that we
work in this paper, the fluid is a perfect one and the brane equations
are the Euler equations for the fluid ---{\it intrinsic} equations---
plus a generalization of the geodesic equation for the motion of a
$p$-brane ---{\it extrinsic} equations for the worldvolume embedding.
We believe that in principle it should be possible to incorporate 
higher-derivative corrections to these and compute transport coefficients 
by performing a derivative expansion of the underlying microscopic theory, 
in this case Einstein's theory. 

Closely related precedents of a mapping of black hole dynamics to fluid
dynamics are the `membrane paradigm' \cite{membrane}, and the more recent
`fluid/AdS-gravity correspondence' \cite{Bhattacharyya:2008jc}. As we shall argue near
the end, the fluid/AdS-gravity correspondence can be embedded within
the approach we advocate here --- in fact it has been an
important influence in developing it. The general arguments discussed
above indicate that a fluid-dynamical description should indeed be
expected to exist for any long-wavelength fluctuations around
an equilibrium state. From this perspective, perhaps the main
qualitative novelty of our approach is that the existence of a hierarchy
of scales in higher-dimensional black holes makes this effective theory
useful not only for studying fluctuations, but also for constructing and
analyzing in a very general manner novel kinds of stationary (\ie\
equilibrium) black holes, including vacuum solutions.

The outline of the paper is the following: Section~\ref{sec:effwv}
develops the conceptual basis underlying the blackfold approach as a
worldvolume theory of the dynamics of black branes.
Section~\ref{sec:bfdyn} presents a main result of this paper:
the {\it blackfold equations}, a set of coupled non-linear differential
equations for the collective coordinates of a neutral black brane.
Section~\ref{sec:stationary} focuses on the important case of stationary
blackfolds, for which the intrinsic subset of these equations can be
explicitly solved. Section~\ref{sec:bfboundaries} analyzes the issues
raised by the possible presence of boundaries of the blackfold
worldvolume. In section~\ref{sec:areamassspin} we describe how to
compute the physical magnitudes of a blackfold.
Section~\ref{sec:actandfirst} presents an action principle for
stationary blackfolds. This is useful for practical calculations, but 
also admits a simple and appealing interpretation in terms of black hole
thermodynamics. Section~\ref{sec:GLinBF} discusses briefly the stability of
blackfolds exhibiting how the approach can uncover in a
remarkably simple way the Gregory-Laflamme instability of black branes.
We close in Section~\ref{sec:discussion} with a discussion of the
relation of blackfolds to other effective theories of black hole
dynamics, in particular the fluid/AdS-gravity correspondence. In the
appendix we collect a number of technical results on the extrinsic
geometry of submanifold embeddings.

\bigskip

\noindent\textit{Notation and terminology:}

For clarity and later reference, we summarize here some of our notation.

For a blackfold of $p$ spatial dimensions in $D$-dimensional spacetime
it is convenient to introduce
\beq
n=D-p-3\,.
\eeq
The codimension of the blackfold worldvolume is $n+2$.

Spacetime (background) and worldvolume magnitudes are denoted and
distinguished as follows:
\begin{itemize}
\item
Spacetime coordinates (and embedding functions): 
$X^\mu$, $\mu,\nu\ldots=0,\dots, D-1$.\\
Background metric: $g_{\mu\nu}$.\\ 
Background metric connection: $\Gamma_{\mu\nu}^\sigma$.\\ 
Background covariant derivative: $\nabla_\mu$.

\item 
Worldvolume coordinates: $\sigma^a$, $a,b\ldots =0,\dots,p$.\\
(Induced) worldvolume metric: $\gamma_{ab}$.\\
Worldvolume metric connection: $\left\{ {a\atop b\,c}\right\}$.\\
Worldvolume covariant derivative: $D_a$.


\end{itemize}

Indices $\mu,\nu,\dots$ are lowered and raised with $g_{\mu\nu}$,
indices $a,b,\dots$ with $\gamma_{ab}$.

We use the same letter for a background tensor tangent to
the worldvolume,
${t_{\mu\dots}}^{\nu\dots}$, and
for its pullback onto the worldvolume, ${t_{a\dots}}^{b\dots}$ (the only
exception is the first fundamental form $h_{\mu\nu}$ and the
induced metric $\gamma_{ab}$).

$\Omega_{(n)}$ denotes the volume of the unit $n$-sphere. $\Omega_i$
denotes the angular velocity of the blackfold in the $i$-th direction.

$V_{(p)}$ is the volume of a spatial section of the blackfold. $V_i$ is the
spatial velocity field on the worldvolume, and $V=\sqrt{\sum_i V_i^2}$.

Note that we refer to `long-wavelength' and not `low-energy', effective
theory, the reason being that in classical gravity, without
$\hbar$, such notions are not equivalent (indeed large energies
typically imply long distances in classical gravity).

\section{Effective worldvolume theory}
\label{sec:effwv}

We present the effective theory of blackfolds trying to highlight the
similarities with the field-theoretical effective description of other
extended objects, such as cosmic strings or D-branes. The main
differences with these are, first, that the short-distance degrees of
freedom that are integrated out are not those of an Abelian Higgs model
nor massive string modes, but rather purely gravitational degrees of
freedom. Second, the extended objects ---curved black branes--- possess
black hole horizons. We obtain the equations using general symmetry and
conservation considerations, rather than doing a detailed derivation
from first principles. 

\subsection{Collective coordinates for a black brane}
\label{sec:collective}

Schematically, the degrees of freedom of General
Relativity are split into long and short
wavelength components,
\beq\label{gsplit}
g_{\mu\nu}=\{g_{\mu\nu}^{(\mathrm{long})},g_{\mu\nu}^{(\mathrm{short})}\}\,.
\eeq 
The Einstein-Hilbert action is then approximated as\footnote{For clarity
of presentation, at this initial stage we consider that both long and
short degrees of freedom obey vacuum gravity dynamics, $R_{\mu\nu}=0$, but
this can be easily generalized, see sec.~\ref{sec:nonvac} below.}
\beq\label{Isplit}
I_{EH}=\frac{1}{16\pi G}\int d^Dx\sqrt{-g}R\approx
\frac{1}{16\pi G}\int d^Dx\sqrt{-g^{(\mathrm{long})}}R^{(\mathrm{long})}+
I_{\mathrm{eff}}[g_{\mu\nu}^{(\mathrm{long})},\phi]\,,
\eeq
where $I_{\mathrm{eff}}[g_{\mu\nu}^{(\mathrm{long})},\phi]$ is an
effective action obtained after
integrating-out the short-wavelength gravitational degrees of
freedom (precisely what we mean by this will be made clear in
sec.~\ref{sec:efftensor}). The coupling of these to the
long-wavelength component of
the gravitational
field is captured through a set of `collective coordinates' that we
denote schematically by $\phi$. Our first task is
to identify these effective field variables and the length scales that
allow this splitting of degrees of freedom.

The main clue to the nature of the effective theory comes from the
observation that the limit
$\ell_M/\ell_J\to 0$ of known black holes, when it exists, results in
flat black branes. Thus we shall take the effective theory to describe
the collective dynamics of
a black $p$-brane. Its geometry in $D=3+p+n$
spacetime dimensions is 
\beq\label{pbrane}
ds^2_{p\mathrm{-brane}}=
-\left(1-\frac{r_0^n}{r^n}\right)dt^2+
\sum_{i=1}^p ({dz^i})^2+\frac{dr^2}{1-\frac{r_0^n}{r^n}}+r^2 d\Omega^2_{n+1}
\,.
\eeq
The coordinates $\sigma^a=(t,z^i)$ span the brane worldvolume. A
more general form of the metric is obtained by boosting it along the
worldvolume. If the velocity field is $u^a$, with $u^a u^b\eta_{ab}=-1$ then
\beq\label{boostpbrane}
ds^2_{p\mathrm{-brane}}=
\left(\eta_{ab}+\frac{r_0^n}{r^n}u_a
u_b\right)d\sigma^a d\sigma^b+\frac{dr^2}{1-\frac{r_0^n}{r^n}}+r^2 d\Omega^2_{n+1}
\,.
\eeq

The parameters of this black brane solution consist of the
`horizon thickness' $r_0$, the $p$ independent components of the
velocity $u$ (say, its spatial components $u^i$), and 
the $D-p-1$ coordinates that parametrize the position of the brane in
directions transverse to the worldvolume, which we denote
collectively by $X^\perp$. 
The $D$ collective coordinates of the black brane are
\beq\label{collcoords}
\phi(\sigma^a)=\{ X^\perp(\sigma^a), r_0(\sigma^a),
u^i(\sigma^a)\}\,
\eeq 
and in the long-wavelength effective theory one allows $\partial X^\perp$,
$\ln r_0$ and $u_i$ to vary slowly along the worldvolume,
$\mathcal{W}_{p+1}$, over a
length scale $R$ much longer than the size-scale of
the black brane, 
\beq
R\gg r_0\,. 
\eeq
Typically the scale $R$ is set by the smallest intrinsic or extrinsic
curvature radius of the worldvolume. Observe that we require slow variations of 
$\partial X^\perp$, not of $X^\perp$. Like the longitudinal
velocities $u^a$, the transverse `velocities' $\partial X^\perp$ can
be arbitrary.

In order to preserve manifest
diffeomorphism invariance it is convenient to introduce some gauge
redundancy and enlarge the set of embedding coordinates of the
worldvolume of the black brane to include all the spacetime coordinates
$X^\mu(\sigma^a)$. 
From this embedding we can compute an induced metric
\beq\label{gammaalbe}
\gamma_{ab}=g_{\mu\nu}^{(\mathrm{long})}\partial_a X^\mu \partial_b X^\nu\,.
\eeq
This is naturally interpreted as the geometry induced on the worldvolume
of the brane. To understand what this means, regard the split between
degrees of freedom as follows: the
long-wavelength degrees of freedom live in a `far-zone' $r\gg r_0$,
and they describe the background geometry in which the (thin) brane
lives. Then \eqref{gammaalbe} is the metric induced on the brane
worldvolume.
The short-wavelength degrees of freedom live in the
`near-zone' $r\ll R$. In the strict limit where $R\to\infty$, the near-zone
solution is \eqref{boostpbrane}, but when $R$ is large but finite, the
collective coordinates depend on $\sigma$. Also, the
long and short degrees of freedom
interact together
in the `overlap' or `matching-zone' $r_0\ll r\ll R$, where the metrics
$g_{\mu\nu}^{(\mathrm{long})}$ and $g_{\mu\nu}^{(\mathrm{short})}$
must match.
Then the near-zone metric for the black brane must be of the form
\beq\label{bbold}
ds^2_{\mathrm{(short)}}=
\left(\gamma_{ab}(\sigma)+\frac{r_0^n(\sigma)}{r^n}u_a(\sigma)
u_b(\sigma)\right)d\sigma^a
d\sigma^b+\frac{dr^2}{1-\frac{r_0^n(\sigma)}{r^n}}+r^2 d\Omega^2_{n+1}
+\dots \,.
\eeq
The dots here indicate that, without additional terms, in general this
is not a solution to the Einstein equations. These equations 
contain terms with gradients of $\ln r_0$, $u^a$ and
$\gamma_{ab}$. However these terms can be seen to come multiplied by powers of
$r_0$ so they are small when $r_0/R\ll 1$. Then we
can consider an expansion of the equations in derivatives and add
a correction to \eqref{bbold}
to find a solution to the Einstein equations to first order in the
derivative expansion. A subset of the resulting Einstein equations can
be rewritten as equations on the collective field variables
$\phi(\sigma)$. An important requirement is that the perturbations
preserve the regularity of the horizon, and to this effect working in a
set of coordinates (Eddington-Finkelstein type) different than the ones
above may be more appropriate. 

The development of this line of argument, which can be regarded as a
blend of the ideas for the effective descriptions of black hole dynamics
in \cite{Poisson:2003nc,Gralla:2008fg} (and references therein), and in
\cite{Bhattacharyya:2008jc}, produces a systematic derivation of the
blackfold equations. This is however a technically involved approach
that we hope to discuss elsewhere. Here we shall instead follow a less
rigorous but quicker and physically well-motivated path, relying on
general effective-theory-type of arguments that allow us to readily
obtain the blackfold formalism valid to lowest order in the derivative
expansion. As we will see, this is the `perfect fluid' and `generalized
geodesic' approximation. The more systematic method outlined above would
be needed to go beyond these approximations and account for dissipation
and effects of internal structure and gravitational 
self-force\footnote{Also, see \cite{cleft}
for an approach to the calculation of these corrections (for
zero-branes) in an effective-theory framework somewhat akin to the
spirit in this paper.}.

We have considered a black brane in \eqref{pbrane}, \eqref{boostpbrane}
that is not rotating along the transverse $(n+1)$-sphere, nor have we
included any possible deformations of it. This is just a simplification
and is not essential. Note first that ultraspins in this sphere, with
rotation parameter $a\propto J/M\gg r_0$, can indeed be considered in
the blackfold approach, but then the starting point must be the black
brane limit that results when $a/r_0\to \infty$. On the other hand small
spins, with $a \lesssim r_0$, can also be included easily to the order
that we work in this paper. The reason is that the modifications to the
blackfold dynamics introduced by these spins only enter at a higher
order in the expansion in $r_0/R$. This is familiar, for instance, in
that spin effects on the worldline of a test particle enter through
couplings to the background curvature tensor: they reflect the internal
structure of the particle, which clearly is a higher-order correction.
Thus the dynamics associated to internal spin and polarization effects
of the spheres of size $r_0$ is effectively integrated-over without
affecting the lowest-order formalism. It must be noted, though, that
higher-dimensional neutral black holes exhibit zero-mode deformations at
discrete values of the spin when $n\geq 3$ (when $n=1,2$ such
deformations appear to be always massive modes) \cite{Dias:2009iu}. For these spins,
these deformations can be excited at arbitrarily low frequencies and
therefore must be added to the collective coordinates of the black
brane. Other than this, the internal spin can be treated just like a
conserved charge in the worldvolume. As such, it must satisfy a
continuity equation but, since the gravitational effects of this spin
fall off faster at large $r$ than those of the mass, the computation of
the effective worldvolume stress tensor in the next subsection is
unaffected. The blackfold equations also apply in the form below in the
presence of internal spins.

\subsection{Effective stress tensor}
\label{sec:efftensor}

By the phrase `integrating out the short-distance dynamics' we mean that
the Einstein equations are solved at distances $r\ll R$ and then the
effects of the solution at distances $r\gg r_0$ are encoded in a
stress-energy tensor that depends only on the collective coordinates.
The stress tensor is such that its effect on the long-wavelength field
$g_{\mu\nu}^{(\mathrm{long})}$ is the same as that of the black brane at
distances $r\gg r_0$. For
reasons that will become apparent as we proceed, it is both simpler and
more convenient to work with an effective stress-energy tensor rather
than with an effective action. In any case, nothing is lost since we
work at the classical level. 

The effective equations from \eqref{Isplit} are
\beq\label{effEins}
R^{(\mathrm{long})}_{\mu\nu}-\frac{1}{2}R^{(\mathrm{long})}
g_{\mu\nu}^{(\mathrm{long})}=
8\pi G T^{\mathrm{eff}}_{\mu\nu}\,,
\eeq
where the effective worldvolume stress tensor is
\beq\label{Teff}
T^{\mathrm{eff}}_{\mu\nu}=-\frac{2}{\sqrt{-g_{(\mathrm{long})}}}\frac{\delta
I_{\mathrm{eff}}}{\delta g_{(\mathrm{long})}^{\mu\nu}}\Bigg\vert_{\WW_{p+1}}\,.
\eeq

We now argue that the appropriate notion for this effective
stress-tensor that captures the coupling of the short-wavelength degrees
of freedom to the long-wavelength ones, is the quasilocal stress-energy
tensor introduced by Brown and York \cite{Brown:1992br}. This is defined
by considering a timelike hypersurface that lies away from the black
brane and encloses it by extending along the worldvolume directions and
the angular directions $\Omega_{(n+1)}$, \ie the hypersurface acts as a
boundary. Actually, as we explained above, the angular directions are
integrated over in our description (and to leading order they do not
play any role), so we can simplify the discussion by focusing
exclusively on the worldvolume directions of the boundary. If the
boundary metric (along worldvolume directions) is $\gamma_{ab}$ then the
quasilocal stress tensor is
\beq\label{Tql}
T_{ab}^{\mathrm{(quasilocal)}}=-
\frac{2}{\sqrt{-\gamma}}\frac{\delta I_{cl}}{\delta \gamma^{ab}}\,,
\eeq
where $I_{cl}$ is the classical on-shell action of the solution. For our
purposes, this is the action where the short-distance gravitational
degrees of freedom, $r\ll R$, are integrated and so it must be the same
function of the collective variables as
$I_{\mathrm{eff}}$. Together with the relation \eqref{gammaalbe}
this implies that we can identify \eqref{Teff} with \eqref{Tql}.

It is shown in \cite{Brown:1992br} that the Einstein equations with an
index orthogonal to the
boundary are first-order equations equivalent to the equation of
conservation of the
quasilocal
stress tensor,
\beq\label{DT}
D_a T^{ab}_{\mathrm{(quasilocal)}}=0\,,
\eeq
where $D_a$ is the covariant derivative associated to the boundary
metric $\gamma_{ab}$. Hence, solving the equations \eqref{DT} is
equivalent to solving (a subset of) the Einstein equations. 

Since we identify the stress tensors \eqref{Teff} and \eqref{Tql},
henceforth we drop the superscripts from them. We also drop the
superscript ${}^{(\mathrm{long})}$ from the background metric
$g_{\mu\nu}$.

The effective stress tensor is computed in the zone $r_0\ll r\ll R$,
where the gravitational field is weak and the quasilocal stress tensor
$T^{ab}$ is, to leading order in $r_0/R$, the same as the ADM stress
tensor. For the boosted black $p$-brane \eqref{boostpbrane} one can
readily compute it and find
\beq\label{blackTab}
T^{ab}=\frac{\Omega_{(n+1)}}{16\pi G}r_0^n\left( n u^a u^b- \eta^{ab}
\right)\,.
\eeq
After introducing a slow variation of the collective coordinates the
stress tensor becomes
\beq\label{blackTab2}
T^{ab}(\sigma)=\frac{\Omega_{(n+1)}}{16\pi G}r_0^n(\sigma)\left( n
u^a(\sigma) u^b(\sigma)- \gamma^{ab}(\sigma)
\right)+\dots
\eeq
where the dots stand for terms with gradients of $\ln r_0$, $u^a$, and
$\gamma_{ab}$, which we are taking to be small and are neglected in this
paper.

\subsection{General branes: fluid perspective}
\label{sec:generalfluids}

On general grounds, the long-wavelength effective theory for any kind of
brane will take the form of a derivative expansion for an effective
stress-energy tensor that satisfies the conservation equations
\eqref{DT}. This is the dynamics of an effective fluid that lives on the
worldvolume spanned by the brane. If the worldvolume theory is 
isotropic, then to lowest derivative order the stress tensor is that of
an isotropic perfect fluid,
\beq\label{perfluid}
T^{ab}=(\vep+P) u^a u^b +P \gamma^{ab}\,,
\eeq
with energy density $\vep$, pressure $P$ and velocity $u^a$ satisfying
\beq
u^a u^b \gamma_{ab}=-1\,.
\eeq
Thermodynamics provides
the universal macroscopic description of equilibrium configurations, and
fluid dynamics is the general long-wavelength description of
fluctuations under the assumption of local
equilibrium. So
in general there will be an equation of state, which we write in
the
form $P(\vep)$, and the system
will obey locally the laws of thermodynamics
\beq\label{firstlaw}
d\vep =\mathcal{T}ds
\eeq
and Euler-Gibbs-Duhem relation
\beq\label{duhem}
\vep+P =\mathcal{T}s
\eeq
where $\mathcal{T}$ is the local temperature and $s$ the entropy density
of the fluid in its rest frame. The fluid may also carry additional
conserved charges, but we do not consider these in this paper.

For a black brane, \eqref{blackTab2} tells us that the effective fluid has
\beq\label{blackepsP}
\vep =\frac{\Omega_{(n+1)}}{16\pi G}(n+1)r_0^n\,,\qquad
P=-\frac{1}{n+1}\,\vep\,.
\eeq
Moreover, in the rest frame of the fluid the Bekenstein-Hawking
identification between horizon area and
entropy
\beq\label{locs}
s=\frac{\Omega_{(n+1)}}{4 G} r_0^{n+1}
\eeq
and between surface gravity and temperature 
\beq\label{locT}
\mathcal{T}=\frac{n}{4\pi r_0}
\eeq
is well known to reproduce the correct thermodynamic relations
\eqref{firstlaw}, \eqref{duhem}.

Going beyond the perfect fluid approximation \eqref{perfluid}, the
stress tensor will acquire dissipative terms
proportional to gradients of $\ln r_0$, $u^a$, $\gamma^{ab}$. As
discussed above, these are neglected in this paper. At any rate their
effects are absent for stationary
configurations.

\section{Blackfold dynamics}
\label{sec:bfdyn}

We have argued that the general effective theory of classical brane
dynamics can be formulated as a theory of a fluid on a dynamical
worldvolume. The fluid variables must satisfy the intrinsic equations
\eqref{DT}, and they will be coupled to the `extrinsic' equations for
the dynamics of the worldvolume geometry, which we still have to
determine. To this effect, in the next subsection we introduce a few
notions about the geometry of worldvolume embeddings. More details and
proofs are provided in the appendix.

\subsection{Embedding and worldvolume geometry}
\label{sec:geometry}

Given the induced metric on $\mathcal{W}_{p+1}$, \eqref{gammaalbe}, the
first fundamental form of the submanifold is
\beq
h^{\mu\nu}=\partial_a X^\mu \partial_b X^\nu \gamma^{ab}\,.
\eeq
Indices $\mu\,,\nu$ are raised and lowered with $g_{\mu\nu}$, and $a,b$
with $\gamma_{ab}$. Defining
\beq
\perp_{\mu\nu}=g_{\mu\nu}-h_{\mu\nu}
\eeq
it is easy to see that the tensor ${h^\mu}_\nu$ acts as a projector
onto $\mathcal{W}_{p+1}$, and ${\perp^\mu}_\nu$ along directions
orthogonal to $\mathcal{W}_{p+1}$. 

Background tensors ${t^{\mu\dots}}_{\nu\dots}$ with support on
$\mathcal{W}_{p+1}$ can be converted into worldvolume tensors
${t^{a\dots}}_{b\dots}$ and viceversa using $\partial_a
X^\mu$. For instance, the velocity field
\beq
u^\mu=\partial_a X^\mu\, u^a\,,
\eeq
preserves its negative-unit norm under this mapping. 

The covariant differentiation of
tensors that live in the worldvolume is well defined only along
tangential directions, which we denote by an overbar,
\beq
\oln_\mu={h_\mu}^\nu\nabla_\nu\,.
\eeq 

Note that in general $\oln_\rho {t^{\mu \dots}}_{\nu\dots}$ has both
orthogonal and tangential components. The tangentially projected
part is essentially the same as the worldvolume covariant derivative
$D_c {t^{a \dots}}_{b\dots}$ for the metric $\gamma_{ab}$, both
tensors being related via the
pull-back map $\partial_a X^\mu$. In particular, the divergence of the
stress-energy tensor
\beq
T^{\mu\nu}=\partial_a X^\mu\partial_b X^\nu\,  T^{ab}\,
\eeq 
satisfies (see \eqref{divs})
\beq\label{divT}
{h^\rho}_\nu \oln_\mu T^{\mu\nu}=\partial_b X^\rho D_a T^{ab}\,.
\eeq

The \textit{extrinsic curvature tensor}
\beq\label{Kext}
{K_{\mu\nu}}^\rho={h_\mu}^\sigma \oln_\nu {h_\sigma}^\rho
\eeq
is tangent to $\mathcal{W}_{p+1}$ along its (symmetric) lower indices
$\mu$, $\nu$, and orthogonal to $\mathcal{W}_{p+1}$ along $\rho$. Its
trace is
the \textit{mean curvature vector}
\beq
K^\rho=h^{\mu\nu}{K_{\mu\nu}}^\rho =\oln_\mu h^{\mu\rho}\,.
\eeq
Explicit expressions for the extrinsic curvature tensor in terms of the
embedding functions
$X^\mu(\sigma^a)$ can be found in the appendix.

\subsection{Blackfold equations}

The general extrinsic dynamics of a brane has been analyzed by Carter in
\cite{Carter:2000wv}. The equations are formulated in terms of a
stress-energy tensor with support on the $p+1$-dimensional worldvolume
$\WW_{p+1}$ satisfying the tangentiality condition
\beq
{\perp^\rho}_\mu T^{\mu\nu}=0\,.
\eeq

The basic assumptions are that (i) this effective
stress-energy tensor derives from an underlying conservative dynamics
(in our case, General Relativity), even if the macroscopic
(=long-wavelength) dynamics may be dissipative; and that (ii) spacetime
diffeomorphism invariance holds, or equivalently, the worldvolume theory
can be consistently coupled to the long-wavelength gravitational field
$g_{\mu\nu}$. 
Under these assumptions, the stress tensor must obey the conservation
equations
\beq\label{nablatmunu}
\oln_\mu T^{\mu\rho}=0\,.
\eeq
These are in fact the generic equations of motion for the entire set of
worldvolume field variables $\phi(\sigma^a)$, both intrinsic and
extrinsic: we can decompose
\eqref{nablatmunu} along directions parallel and orthogonal to
$\WW_{p+1}$ as
\beqa
\oln_\mu T^{\mu\rho}&=&\oln_\mu(T^{\mu\nu}{h_\nu}^\rho)=
T^{\mu\nu}\oln_\mu{h_\nu}^\rho+{h_\nu}^\rho\oln_\mu T^{\mu\nu}\nonumber\\
&&=T^{\mu\nu}{h_\nu}^\sigma\oln_\mu{h_\sigma}^\rho+{h_\nu}^\rho\oln_\mu T^{\mu\nu}\nonumber\\
&&= T^{\mu\nu} {K_{\mu\nu}}^\rho +\partial_b X^\rho D_a T^{ab}
\eeqa
where in the last line we used \eqref{divT} and \eqref{Kext}.
Thus the $D$ equations \eqref{nablatmunu} separate into $D-p-1$ equations
in directions orthogonal to $\WW_{p+1}$ and $p+1$ equations parallel
to $\WW_{p+1}$,
\beqa
T^{\mu\nu} {K_{\mu\nu}}^\rho&=&0 \qquad 
\mathit{(extrinsic\ equations)}\,,\label{extreqs}\\
D_a T^{ab}&=&0 \qquad \mathit{(intrinsic\ equations)}\,.\label{intreqs}
\eeqa

Let us now apply the equations \eqref{nablatmunu} onto the
generic stress-energy
tensor of a perfect fluid on the
worldvolume
\beq
T^{\mu\nu}=(\vep +P)u^\mu u^\nu+P h^{\mu\nu}\,.
\eeq
We find
\beq\label{geneqs}
u^\mu u^\nu\oln_\nu\vep +(\vep+P)(\dot u^\mu +u^\mu\oln_\nu u^\nu)
+(h^{\mu\nu}+u^\mu u^\nu)\oln_\nu P+P K^\mu=0\,,
\eeq
where 
\beq
\dot u=u^\nu\nabla_\nu u\,
\eeq
is the acceleration of $u^\mu$. These are the general equations, to
leading order in the derivative expansion, for the dynamics of a
classical brane with worldvolume spatial isotropy, possibly supplemented
by conservation equations for charges, if present. A familiar example is a
Nambu-Goto-Dirac brane, with $T_{\mu\nu}=-|P|h_{\mu\nu}$ and $\nabla
P=\nabla \vep=\nabla u=0$, for which the extrinsic equations,
$K^\rho=0$, require that the worldvolume be a minimal submanifold. But
any classical brane will satisfy
equations of this form.

As usual the projection of \eqref{geneqs}
onto $u$ is the continuity equation for the energy of
the fluid,
\beq\label{inteq1}
u^\nu\oln_\nu\vep+(\vep+P)\oln_\mu u^\mu=0\,,
\eeq
while the projections orthogonal to $u$
\beqa
(\vep +P)\dot u^\mu
=-(h^{\mu\nu}+u^\mu u^\nu)\oln_\nu P
-P K^\mu
\eeqa
say that the force that accelerates an element of the
fluid is given along worldvolume directions by pressure gradients
(Euler equation)
and in directions transverse to the worldvolume by the extrinsic
curvature.

For the specific stress tensor of a neutral black brane,
\eqref{blackepsP}, the equations \eqref{geneqs} become, after a little
manipulation,
\beq\label{bfeqs}
\dot u^\mu+\frac{1}{n+1}u^\mu \oln_\nu u^\nu =\frac{1}{n}K^\mu+\oln^\mu\ln
r_0\,
.
\eeq
These \textit{blackfold equations} describe the general collective dynamics of
a neutral black brane.

Again, we can
decompose them into different projections. In directions orthogonal to
the worldvolume we have
\beq\label{bfeq1}
K^\rho=n{\perp^\rho}_\mu \dot u^\mu\,.
\eeq
The equivalence of this equation to \eqref{extreqs} follows by using
\eqref{vvK}. 

The equations parallel to the worldvolume are
\beq
h_{\mu\nu}\dot u^\nu +\frac{1}{n+1}u_\mu \oln_\nu u^\nu =\oln_\mu\ln
r_0\,,
\eeq
which we can also write using worldvolume indices and derivatives,
\beq\label{bfeq2}
\dot u_a+\frac{1}{n+1}u_a D_b u^b =\partial_a\ln r_0\,,
\eeq
with $\dot u^b=u^c D_c u^b$.
Thus the temporal and spatial worldvolume gradients of $r_0$ determine the
worldvolume acceleration and expansion of $u$, respectively.

Although we have emphasized the fluid-dynamical interpretation of the
equations, it is interesting to observe that the extrinsic equations
\eqref{extreqs}, when written explicitly in terms of the embedding
$X^\mu(\sigma^a)$ become
\beq
T^{ab}{\perp_\sigma}^\rho\left(\partial_a\partial_b X^\sigma
+\Gamma^{\sigma}_{\mu\nu}\partial_a X^\mu\partial_b
X^\nu\right)=0\,,
\eeq
or alternatively
\begin{equation}
\label{embextr}
T^{ab}\left(D_a
\partial_b X^\rho + \Gamma^\rho_{\mu \nu} \partial_a X^\mu
\partial_b X^\nu\right)=0\,
\end{equation}
(see eqs.~\eqref{niceextr1} and \eqref{niceextr}).
These can be regarded as generalizations to $p$-branes of the geodesic
equation for free particles, or more simply, of
``mass$\times$acceleration$=0$".

Blackfolds differ from other branes in that they represent objects with
black hole horizons. In the long-distance effective theory we lose sight
of the horizon, since its thickness is of the order of the scale $r_0$
that we integrate out. But the presence of the horizon is reflected in
the effective theory in the existence of an entropy and in the local
thermodynamic equilibrium of the effective fluid. Indeed, we shall
\textit{assume} that the regularity of the event horizon under
long-wavelength perturbations ---including those that bend the
worldvolume away from the flat geometry or that excite the effective
fluid away from equilibrium--- is satisfied when the blackfold
equations, which incorporate in particular local thermodynamic
equilibrium, are satisfied. A proof of this statement requires the
rigorous derivation, outlined at the end of sec.~\ref{sec:collective},
of the blackfold equations in the derivative expansion of Einstein's
equations, which is outside the scope of this paper. However, there is
already significant evidence that horizon regularity is preserved for
blackfold solutions. First, analyses of the perturbations of black
strings that bend them into a circle \cite{Emparan:2007wm} (and
extensions thereof to branes curved into tori) show that the extrinsic
equations \eqref{extreqs} are equivalent to demanding absence of
singularities on or outside the horizon. Second, the intrinsic,
hydrodynamical perturbations of a black brane in AdS have been studied
in detail in \cite{Bhattacharyya:2008jc} and shown to be consistent with
horizon regularity. Note however that the regular solution to higher
orders may not preserve the same symmetries as the lowest-order
solution. In particular, in some cases horizon regularity may require to
abandon stationarity of the configuration at higher orders. In most
instances it appears easy to decide from the physics of the problem
whether such an effect is expected (specific examples will be discussed
elsewhere), but it would be good if precise conditions could be stated
in generality.\footnote{The example studied in \cite{LeWitt:2009qx}
appears to fall outside the remit of our approach since the blackfold (a
small black hole) is not a small perturbation of the background
spacetime.}

\subsection{The metric at all length scales: Matched asymptotic
expansion}

Under the splitting in \eqref{Isplit}, the set of field variables in the
system are the collective worldvolume fields, intrinsic and extrinsic,
and the background gravitational field $g_{\mu\nu}$. The complete set of
equations are the extrinsic equations \eqref{extreqs}, intrinsic
equations \eqref{intreqs}, and backreaction equations \eqref{effEins}.
Since they are a consequence of general symmetry and conservation
principles, these equations retain their form at any perturbative
order.\footnote{This, however, is a somewhat formal statement due to the
appearance of gravitational self-force divergences on the worldvolume
that must be dealt with carefully \cite{Poisson:2003nc}.
Ref.~\cite{Mino:1996nk} shows how the equation of stress tensor
conservation can be used as the basis to obtain these corrections to
particle motion.} The specific form of the stress tensor, as well as the
background metric, will in general be corrected at higher orders.

The only equations that one has to solve at the lowest order are those
that suffice to ensure that $T_{\mu\nu}$ can be consistently coupled to
the long-wavelength gravitational field. These are just the intrinsic
and extrinsic equations, and backreaction is neglected. The explicit
blackfold equations \eqref{bfeqs} that result are valid only for test
branes.

From the point of view of effective field theory one is interested only
in quantities that are measured in the long-wavelength regime. The
short-wavelength dynamics enters only to determine the coefficients in
the effective stress tensor, which can be computed by, \eg matching the
calculations of some observables \cite{cleft}. However, in General
Relativity one is often interested in also having an explicit solution,
even if an approximate one, for the geometry at all scales, including
near the horizon of the black hole. The systematic way to construct an
explicit metric in an expansion in $r_0/R$ is through the method of
matched asymptotic expansions \cite{Poisson:2003nc}. In the context of
blackfolds this was discussed in \cite{Emparan:2007wm} (following
\cite{Harmark:2003yz,Gorbonos:2004uc}), and explicitly applied to the
construction of higher-dimensional black rings. We review it briefly
here. 

As described in sec.~\ref{sec:collective}, the full geometry splits into
near- and far-zones, that share a common overlap-zone. To zeroth order
in $r_0/R$ the near-zone metric is \eqref{boostpbrane} and the far-zone
metric is the background metric $g_{\mu\nu}$. This is as far as we go in
this paper in terms of providing explicit solutions to the Einstein
equations: this is the test-brane approximation. The next order involves
the gravitational backreaction of the brane: the equations
\eqref{effEins} are linearized around the background and solved with the
distributional worldvolume source $T_{\mu\nu}$. The blackfold equations
would appear here among the Einstein equations as constraints
(first-order equations), but we can assume they have already been solved
at zeroth order. The solution of the linearized equations with
appropriate asymptotics (\eg asymptotic flatness) produces a corrected
far-zone metric with corrections of order $(r_0/R)^n$. Its value in the
overlap-zone provides new asymptotic conditions for the near-zone
solution. The next step is to perturb the metric \eqref{boostpbrane}
linearly, with the boundary conditions that the horizon remains regular
and that in the overlap zone the metric matches the corrected far-zone
solution. In this manner we produce a new, corrected solution, at all
scales. This process can then be iterated to higher orders. 

\subsection{Lumpy blackfolds}
\label{sec:lumpy}

The formalism developed in this paper can easily be applied to other
branes once their effective equation of state is known. In particular,
there are other neutral black branes in vacuum gravity than the `smooth'
black branes of \eqref{pbrane}:
refs.~\cite{Gubser:2001ac,Wiseman:2002zc} have shown that `lumpy' black
branes exist, branching off at the threshold of the Gregory-Laflamme
(GL) instability \cite{Gregory:1993vy}. Their horizons are inhomogeneous
on a scale $\sim r_0$, so this small-scale inhomogeneity is averaged
over in our effective description. However, there is an effect on the
effective stress tensor measured at large distance from the brane, since
the equation of state (which is known only perturbatively near the GL
threshold, or possibly numerically) is in general different than the one for
smooth branes \eqref{blackepsP}.

We can use these lumpy branes as the basis for the construction of lumpy
blackfolds. It should be clear that their worldvolume is smooth, but the
horizons of these blackfolds are inhomogeneous on the scale $r_0$. The
simplest example would be a lumpy black ring, built by bending a lumpy
black string into a circular shape. This example serves also to
illustrate a feature of lumpy blackfolds: the lumps in a rotating lumpy
black ring will emit gravitational radiation, so the ring will lose mass
and angular momentum as it evolves in time. In more generality, if the
lumps on a blackfold extend along a direction in which the
fluid velocity is non-zero, and if this direction is not an isometry, then
the lumps moving along these orbits
will give rise to a varying quadrupole and hence to gravitational
radiation. 

Once this effect is taken into account, lumpy blackfolds are generically
expected to evolve in time and not remain stationary. However, the
time-scale for this evolution will be very long. The effect is only
visible when the small scale is resolved, so it is suppressed by a power
of $r_0/R$. It will be further suppressed by the fact that gravitational
radiation couples to a higher-multipole (quadrupole) and therefore is
rather inefficient. In \cite{Elvang:2006dd} this time scale was estimated for
five-dimensional rings, and extending this estimate to black rings in $D=4+n$
dimensions we find the time to be of order $T_{gw}\sim R(R/r_0)^n$, longer
by a factor $(R/r_0)^{n+1}$ than the short time-scale $r_0$. Lumpy
black branes may also be affected by GL-like instabilities, but these
have not been investigated yet.

\subsection{Generalization to non-vacuum theories}
\label{sec:nonvac}

In \eqref{gsplit} and \eqref{Isplit} we assumed that the full dynamics
at all wavelengths is described by vacuum General Relativity, \ie the
Einstein-Hilbert action with no matter nor cosmological constant.
However, this is not actually necessary for our derivation of the
equations of motion and the effective stress tensor. The only part of
the field that is actually required to be governed by the
Einstein-Hilbert action is the sector of short-wavelength degrees of
freedom that we integrate out in order to obtain the effective stress
tensor \eqref{blackTab}. For the long-wavelength components we only
require diffeomorphism invariance, which implies the equations of motion
\eqref{nablatmunu}. Thus, the blackfold equations \eqref{bfeqs} are
enough to describe neutral blackfolds in any configuration that, at small
distances, is dominated by the Einstein-Hilbert term. For instance, this
will be the case for blackfolds in the presence of a cosmological
constant as long as 
\beq
r_0\ll
|\Lambda|^{-1/2}
\eeq
(see \cite{Caldarelli:2008pz} for an explicit application), or for blackfolds in an
external background gauge field as long as the typical
length scale of the background field around the blackfold is much
larger than $r_0$. No restriction on $R$ other than $R\gg r_0$ needs to
be imposed.

A different situation arises for charged blackfolds, since then the gauge
field has short-wavelength components. The effective fluid is then
charged, and additional current conservation equations must be added.
This extension of our analysis will be discussed elsewhere.

\section{Stationary blackfolds}
\label{sec:stationary}

Equilibrium configurations that remain stationary in time are of
particular interest. For a blackfold, they correspond to stationary
black holes. In this case it is possible to solve the blackfold
equations explicitly for the worldvolume variables, namely the thickness
$r_0$ and velocity $u$, so one is left only with the extrinsic equations
for the worldvolume embedding $X^\mu(\sigma)$. 

We employ a general result proven in \cite{Caldarelli:2008mv} for
stationary fluid configurations: if dissipative effects must be absent,
then the fluid
(intrinsic)
equations require that the velocity field be proportional to a
worldvolume Killing field $\mathbf{k}=k^a\partial_a$. That is,
\beq\label{uzeta}
u=\mathbf{k}/|\mathbf{k}|
\eeq
where 
\beq\label{bark}
|\mathbf{k}|=\sqrt{-\gamma_{ab}k^a k^b}
\eeq
and $\mathbf{k}$
satisfies the worldvolume
Killing equation\footnote{For a stationary {\it conformal} fluid this is
relaxed to the conformal Killing equation $D_{(a}k_{b)}=\lambda
\gamma_{ab}$. Conformal fluids do not dissipate the expansion of $u$
since the bulk viscosity vanishes.}
\beq\label{Killwv}
D_{(a}k_{b)}=0\,.
\eeq
This does not necessarily mean
that $\mathbf{k}$ is a Killing field of the background away from
$\WW_{p+1}$, as
the worldvolume could be at a locus of enhanced symmetry. However,
this is not generic, and indeed when the blackfold
thickness is small but non-zero, $\mathbf{k}$ should satisfy the Killing
equations on a finite region around the worldvolume. Thus we assume
the existence of a timelike Killing vector $k^\mu \partial_\mu$ in the
background, 
\beq\label{Killbck}
\nabla_{(\mu}k_{\nu)}=0\,,
\eeq
whose pull-back onto the worldvolume determines the velocity field as in
\eqref{uzeta}. The
existence of a timelike Killing vector field is in
fact a necessary assumption if we intend to describe stationary
black holes. 

The contraction of the Killing equation \eqref{Killbck} with $k^\mu
k^\nu$ implies $k^\mu\partial_\mu
|\mathbf{k}|=0$, and using this it follows easily that\footnote{In
\cite{Caldarelli:2008mv} it is shown that the worldvolume fluid equations directly imply
$u^b D_b u^a=\partial^a \ln |\mathbf{k}|$, which implies the worldvolume
projection of \eqref{unuz}, but not the orthogonal component of it,
which will be used later below in eq.~\eqref{extstat}.}
\beq\label{unuz}
\dot u^\mu=\partial^\mu \ln |\mathbf{k}|\,.
\eeq
Since the expansion of $u$ vanishes, the intrinsic blackfold equation
\eqref{bfeq2} becomes
\beq
\partial_a \ln |\mathbf{k}|=\partial_a\ln r_0\,
\eeq
so
\beq\label{r0zeta}
\frac{r_0}{|\mathbf{k}|}=\mathrm{constant}\,.
\eeq
In order to fix the proportionality constant in \eqref{r0zeta}, we turn
to the fact that the stationary blackfold describes a black brane
with a Killing horizon. 

Above we have defined $\mathbf{k}$ as a Killing vector in the
background. Its norm $\mathbf{k}^2=-|\mathbf{k}|^2$ computed with the worldvolume
metric $\gamma_{ab}$ is negative, \ie\ it is a timelike vector. But
recall that we regard the background metric as only one part, the
far-zone metric at
$r\gg r_0$, of the full
geometry. For $r\ll R$ the geometry is instead well approximated
by the near-zone metric \eqref{bbold}, which matches with the far-zone
metric in $r_0\ll
r\ll R$. It is
natural to extend $\mathbf{k}$ as a Killing vector to all the geometry,
including the near-zone
\eqref{bbold}. Using the metric $g^{\mathrm{(short)}}_{\mu\nu}$ in this region the
norm of 
$\mathbf{k}$ is
\beq
g^{\mathrm{(short)}}_{\mu\nu} k^\mu k^\nu =\left(\gamma_{ab}+\frac{r_0^n}{r^n} 
u_a u_b\right)k^a k^b
=-\left(1-\frac{r_0^n}{r^n}\right)|\mathbf{k}|^2
\eeq
where we used \eqref{uzeta} and \eqref{bark}. Thus
 $\mathbf{k}$ becomes null as we approach the horizon $r\to
r_0$ and indeed it is the
null Killing generator of the horizon. Its surface gravity is
easily obtained as
\beq\label{kapp}
\kappa=\frac{n |\mathbf{k}|}{2 r_0}\,.
\eeq
Equation \eqref{r0zeta} tells us that $\kappa$ is a constant over the
worldvolume of the blackfold. That is, the surface gravity is
uniform not only over the $(n+1)$-sphere but over the entire horizon.

Eqs.~\eqref{uzeta} and \eqref{kapp} provide the general solution to the
intrinsic equations for stationary blackfolds. To give them a more
explicit expression it is convenient to choose a preferred set of 
orthogonal commuting vectors of the background
geometry, $\xi$ timelike and $\chi_i$ spacelike, 
such that the Killing vector $\mathbf{k}$ is a linear
combination of them 
\beq
\mathbf{k} =\xi +\sum_i \Omega_i \chi_i\,,
\eeq
with constant $\Omega_i$. Clearly $i$ runs at most up to $p$. A
convenient (but not necessary) choice is to take for $\xi$ the generator
of asymptotic time translations, and for $\chi_i$ the Cartan generators
of asymptotic rotations with closed orbits of periodicity $2\pi$. They
will often be Killing vectors themselves, but in principle only the
linear combination in $\mathbf{k}$ need be so. 
Introduce a set of worldvolume functions $R_a(\sigma)$ via their
norms on the worldvolume,
\beq\label{Ra}
R_0=\sqrt{-\xi^2}\Big\vert_{\WW_{p+1}}\,,
\qquad R_i=\sqrt{\chi_i^2} \Big\vert_{\WW_{p+1}}\,.
\eeq
The $R_a$ must be regarded as part of the embedding coordinates
$X^\mu(\sigma)$. 
If $\xi$ is the canonically-normalized
generator of asymptotic time translations then $R_0$ is a
redshift factor between infinity and the blackfold worldvolume. The
$R_i$ are the proper radii of the orbits
generated by $\chi_i$ along the worldvolume. The $\Omega_i$ are the
horizon angular velocities relative to observers that follow orbits
of $\xi$. 

The vectors
\beq\label{xichi}
\frac{\partial}{\partial t}=\frac{1}{R_0}\xi\,,\qquad
\frac{\partial}{\partial z^i}= \frac{1}{R_i} \chi_i\,
\eeq
(no sum in $i$) are orthonormal with respect to the metric
$\gamma_{ab}$ on the worldvolume. It will also be useful to regard them
as vectors that extend into all of
\eqref{bbold}. 

Let us introduce the worldvolume spatial velocity field
\beq
V_i(\sigma)=\frac{u\cdot\partial_{z^i}}{-u\cdot
\partial_t}
=\frac{\Omega_i R_i(\sigma)}{R_0(\sigma)}
\eeq
so that
\beq
\mathbf{k}=R_0\left(\frac{\partial}{\partial t}
+\sum_i V_i\frac{\partial}{\partial z^i}
\right)\,
\eeq
and 
\beqa\label{modk}
|\mathbf{k}|&=&\left(-\xi^2-\sum_i \Omega_i^2 \chi_i^2\right)^{1/2}\nonumber\\
&=&R_0\sqrt{1-V^2}\,,
\eeqa
where
\beq
V^2=\sum_i V_i^2=\frac{1}{R_0^2}\sum_i \Omega_i^2 R_i^2\,.
\eeq
Thus $|\mathbf{k}|$ can be regarded as the relativistic Lorentz factor
at a point in $\WW_{p+1}$,
with a possible local redshift, all relative to the reference frame of
$\xi$-static observers.
Plugging \eqref{modk} into \eqref{kapp} we obtain that for given $\kappa$
and $\Omega_i$ the
thickness $r_0$ is solved in terms of the $R_a$ as
\beq\label{r0soln}
r_0(\sigma)=\frac{n R_0(\sigma)}{2\kappa}\sqrt{1-V^2(\sigma)}\,.
\eeq

The conditions that $\kappa$ and $\Omega_i$ must remain uniform over the
blackfold worldvolume were referred to in \cite{Emparan:2009cs} as the
\textit{blackness conditions} and were imposed, invoking general
theorems for stationary black holes --- zeroth law of black hole
mechanics and horizon rigidity
\cite{Hawking:1971vc,Kay:1988mu,Hollands:2006rj} ---, as necessary
conditions for the regularity of the black hole horizon. Here instead we
have derived them as general consequences of stationary fluid dynamics,
where $\kappa$ and $\Omega_i$ appear as integration constants. 

As a matter of fact it is also possible to work entirely within the
framework of the effective theory and avoid any reference to the
short-wavelength geometry of the horizon. Using only the fluid and
thermodynamics equations one can derive, like in
\cite{Caldarelli:2008mv}, that the variation of the local temperature
$\mathcal{T}$ \eqref{locT} along the worldvolume is dictated by the
local redshift
\beq
\mathcal{T}=\frac{T}{R_0\sqrt{1-V^2}}\,.
\eeq
The integration constant $T$ can then be interpreted, using the
thermodynamic first law that we derive below, as the overall temperature
of the black hole. As expected, $T=\kappa/2\pi$. However, we think it is
instructive to see how the concepts of black hole physics, like
null horizon generators and surface gravity, are recovered in this scheme.

\section{Blackfolds with boundaries}
\label{sec:bfboundaries}

Typically branes (such as D-branes or cosmic strings) acquire boundaries
when they intersect or end on other branes. This effect is accounted for by adding
boundary terms to the effective stress tensor. However, black branes
(and other fluid branes) may also have `free' boundaries
without any boundary stresses.

Consider the case where the worldvolume of the brane has a timelike
boundary $\partial\WW_{p+1}$. Assume there is a smooth (but otherwise
arbitrary) extension
$\widetilde\WW_{p+1}$ of $\WW_{p+1}$ across the boundary. Introduce 
a level-set function $f(\sigma^a)$ such that 
$f>0$ in $\WW_{p+1}$ and $f<0$ in $\widetilde\WW_{p+1}-\WW_{p+1}$, 
with $f(\sigma^a)=0$ on the boundary $\partial\WW_{p+1}$. Then
$-\partial_a f$ is a one-form normal to 
$\partial\WW_{p+1}$ pointing away from the fluid.
The stress tensor is
\beq
T_{ab}=\left[(\vep+P)u_a u_b+P\gamma_{ab}\right]\Theta(f)\,,
\eeq
where $\Theta$ is the step function. 

If the fluid is to remain within its bounds then the boundary must be
advected with the fluid, \ie it must be Lie-dragged by $u$,
\beq
\Ld_{u\vert_{\partial\WW_{p+1}}}f=0
\eeq
or equivalently, the velocity must remain parallel
to the boundary,
\beq\label{boundfluid}
 u^a \partial_a f\big\vert_{\partial\WW_{p+1}}=0\,.
\eeq

At the boundary, the stress-energy conservation equation \eqref{intreqs} becomes
\beq
\left((\vep+P)u_a u_b+P\gamma_{ab}\right)\partial^a f\big\vert_{\partial\WW_{p+1}}=0\,.
\eeq
Imposing \eqref{boundfluid} we find that the pressure must approach
zero at the boundary, 
\beq
P\big\vert_{\partial\WW_{p+1}}= 0\,.
\eeq 
This is simply the Young-Laplace equation for a bounded fluid when there is no
surface tension that could balance the fluid pressure at the boundary.
That is, we do not introduce any such boundary stresses, although, as we
mentioned they are of interest when one studies brane intersections.

For a neutral blackfold, vanishing pressure at the boundary means 
\beq\label{bdryr0}
r_0\big\vert_{\partial\WW_{p+1}}= 0\,,
\eeq 
which has a nice geometric interpretation: the thickness of
the horizon must approach zero size at the boundary, so the
horizon closes off at the edge of the blackfold.

If the blackfold is stationary, the condition \eqref{bdryr0} means that
$|\mathbf{k}|\to 0$ so the fluid velocity
becomes null at the boundary. This may happen either because the
boundary is an infinite-redshift surface, $R_0\to 0$, or perhaps more commonly,
because the fluid approaches the speed of light at the
boundary,
\beq
V^2\big\vert_{\partial\WW_{p+1}}= 1\,.
\eeq
An example of this was presented in \cite{Emparan:2009cs}, where a rigidly-rotating
blackfold disk was constructed and shown to reproduce accurately the
properties of MP black holes
with one ultraspin. 

The extrinsic equations \eqref{bfeq1} must also be
satisfied and they impose further
constraints on the worldvolume variables. In the stationary case, using
\eqref{unuz} and \eqref{r0zeta} we see that if the extrinsic
curvature of the worldvolume remains
finite at $\partial \WW_{p+1}$, then not only $r_0\to 0$ but also
$\perp^{\rho\mu}\partial_\mu
r_0$ must vanish there at least as quickly as $r_0$.

\section{Horizon geometry, mass and angular momenta}
\label{sec:areamassspin}

The blackfold construction puts, on any point in the spatial section
$\BB_p$ of $\WW_{p+1}$, a (small) transverse sphere $s^{n+1}$
with Schwarzschild radius
$r_0(\sigma)$. Thus the blackfold represents a black hole with a horizon
geometry that is a product of $\BB_p$ and $s^{n+1}$ --- the product is
warped since the radius of the $s^{n+1}$ varies along $\BB_p$. The null
generators of the horizon are proportional to the velocity field $u$.

If $r_0$ is non-zero everywhere on $\BB_p$
then the $s^{n+1}$ are trivially fibered on $\BB_p$ and the horizon
topology is 
\begin{equation}
\label{embedca}
(\mbox{topology of } \BB_p) \times s^{n+1}\,.
\end{equation}
However, we have seen that if $\BB_p$ has boundaries then $r_0$ will
shrink to zero size at them, resulting in a non-trivial fibration and
different topology. A simple but very relevant instance of this happens when
$\BB_p$ is a topological $p$-ball. Then the horizon topology can easily
be seen to be $S^{p+n+1}=S^{D-2}$.

To analyze the horizon geometry we go to the metric \eqref{boostpbrane}
that locally
describes the geometry of the blackfold to lowest order in $r_0/R$ in
the region $r\ll R$.
There we can choose a local orthonormal frame
$(\partial_t,\partial_{z^i})$ on the
worldvolume, such that $\partial_t$ coincides in the overlap zone
$r_0\ll r\ll R$ with the timelike unit normal $n^\mu$ to $\BB_p$,
\beq\label{normalB}
n^\mu=\left(\partial_t\right)^\mu\,.
\eeq
To lowest order in $r_0/R$ the worldvolume metric is Minkowski and the
spatial
metric on the horizon at $r=r_0$ is 
\beq
ds^2_H=\left(\delta_{ij}+u_i u_j\right)dz^i dz^j+r_0^2 d\Omega_{(n+1)}^2\,
\eeq
with $u_i=u\cdot\partial_{z^i}$. Then the local area density of the
horizon $a_\mathrm{H}$ at a given
point in $\BB_p$ is
\beq
a_\mathrm{H}
=\Omega_{(n+1)}r_0^{n+1}\sqrt{1+\delta_{ij}u_i u_j}\,.
\eeq
For a stationary blackfold the choice for $\partial_t$ and
$\partial_{z^i}$ was made in
\eqref{xichi}, so
\beq\label{normalBs}
n^\mu=\frac{1}{R_0}\xi^\mu\,,
\eeq
and
\beq
u_i=\frac{\Omega_i R_i}{R_0\sqrt{1-V^2}}
\eeq
so the area density is
\beqa
a_\mathrm{H}&=&\frac{\Omega_{(n+1)}r_0^{n+1}}{\sqrt{1-V^2}}\nonumber\\
&=&\Omega_{(n+1)} \left( \frac{n}{2 \kappa} \right)^{n+1} R_0^{n+1}(\sigma^a) 
( 1- V^2(\sigma^a))^{\frac{n}{2}}\,.
\eeqa

The total area of the horizon is then
\beq\label{AH}
A_\mathrm{H}=\int_{\mathcal{B}_p}dV_{(p)}\;a_\mathrm{H}(\sigma^a)\,,
\eeq
where $dV_{(p)}$ denotes the volume form in $\mathcal{B}_p$.

Again, we could have avoided any reference to the geometry
of the horizon and worked instead exclusively with quantities defined
in the effective fluid theory. The entropy density
of the blackfold fluid is given in \eqref{locs} and
after taking the relativistic Lorentz factor $\sqrt{1-V^2}$ into
account, the total entropy is
\beq\label{Sbf}
S=\int_{\mathcal{B}_p}dV_{(p)}\frac{s}{\sqrt{1-V^2}}=\frac{\Omega_{(n+1)}}{4G} \left( \frac{n}{2 \kappa}
\right)^{n+1}\int_{\mathcal{B}_p}dV_{(p)}\;
 R_0^{n+1}(\sigma^a) 
( 1- V^2(\sigma^a))^{\frac{n}{2}}=\frac{A_H}{4G}\,,
\eeq
in agreement with the geometric area computed from \eqref{AH}.
The geometric interpretation involves short-wavelength physics, but is
useful
for exhibiting how the blackfold construction gives precise information
about the entire horizon geometry, including the size of the $s^{n+1}$.

The mass and angular momenta are conjugate to the generators of
asymptotic time translations and rotations, which we assume
are the vectors $\xi$ and $\chi_i$ that we
introduced in sec.~\ref{sec:stationary}.\footnote{In a more general case the
relation may involve linear combinations, but this, although
straightforward, comes
at the expense of more cumbersome expressions.}
Then
\beq\label{MJ}
M= \int_{\mathcal{B}_p}dV_{(p)}\; T_{\mu\nu}n^\mu\xi^\nu\,,\qquad
J_i= -\int_{\mathcal{B}_p}dV_{(p)}\;T_{\mu\nu}n^\mu\chi_{i}^\nu\,.
\eeq
Plugging here \eqref{normalBs} and the results from
sec.~\ref{sec:stationary} we obtain
\beq\label{mass}
M=\frac{\Omega_{(n+1)}}{16 \pi G} \left(\frac{n}{2\kappa}\right)^n
\int_{\mathcal{B}_p}dV_{(p)}\; R_0^{n+1}(1-V^2)^\frac{n-2}{2}\left(n+1-V^2\right)\,,
\eeq
and
\beq\label{angmom}
J_i=\frac{\Omega_{(n+1)}}{16 \pi G} \left(\frac{n}{2\kappa}\right)^n
 n \Omega_i\int_{\mathcal{B}_p}dV_{(p)}\; 
R_0^{n-1} (1-V^2)^\frac{n-2}{2} R_i^2\,.
\eeq

It is easy to extract some interesting consequences of these results.
Let us assume that all length scales along $\mathcal{B}_p$ are $\sim R$
and that the velocities and redshift are moderate (\ie $1-V^2$ and $R_0$ of
order one) over almost all the blackfold. Then the two black hole length
scales introduced in
\eqref{scales} are
\beq
\ell_M\sim (r_0^n R^p)^\frac{1}{D-3}\qquad \ell_J \sim R\,,
\eeq
and the small expansion parameter for the effective theory is
\beq
\left(\frac{\ell_M}{\ell_J}\right)^{D-3}
\sim\left(\frac{r_0}{R}\right)^{n}\,.
\eeq

It is interesting to compare the areas (\ie entropies) of different
blackfolds in a given dimension $D$. In \cite{Emparan:2007wm} we introduced the
dimensionless angular momentum $j$ and dimensionless area $a$ for a
given mass\footnote{One should not confuse the dimensionless total area for
fixed mass $a$ with the blackfold area density $a_\mathrm{H}$.}
\beq
j\sim \frac{J}{M(GM)^{1/(D-3)}}\sim \frac{\ell_J}{\ell_M}\,,\qquad
a\sim \frac{A_H}{\ell_M^{D-2}}\,.
\eeq
The blackfold approximation requires $j\gg 1$. Since $A_H\sim
r_0^{n+1}R^p$ we find that
\begin{equation}
a (j) \sim j^{-\frac{p}{D-3-p}}\,.
\end{equation}
This shows that for a given number of large non-zero angular momenta,
the blackfold with smallest $p$ is entropically preferred at fixed mass.
This is just like we observed for single-spin MP black holes and black
rings in \cite{Emparan:2007wm}: for a given mass, the smaller $p$, the thicker the
horizon, thus the cooler the black hole, and (since $\kappa A_H \sim GM$ \ie
constant for fixed mass) the higher its entropy. 

\section{Action principle and first law of stationary blackfolds}
\label{sec:actandfirst}

\subsection{Action for the embedding geometry}
\label{sec:action}

We have presented in sec.~\ref{sec:stationary} the general solution to the
intrinsic equations for a stationary blackfold. Given a Killing vector
field $\mathbf{k}$, eqs.~\eqref{uzeta},
\eqref{modk} and \eqref{r0soln} allow to eliminate
the variables $r_0(\sigma)$ and $u^a(\sigma)$ in terms of the embedding
functions $X^\mu(\sigma)$. 
The
remaining extrinsic equation \eqref{bfeq1} that determines the embedding
can be written, using \eqref{unuz}, as
\beq\label{extstat}
K^\rho=\perp^{\rho\mu} \partial_\mu \ln |\mathbf{k}|^n\,.
\eeq
Using the result \eqref{varI} from the appendix, this equation can be
obtained from the
action
\beq\label{statact}
I[X^\mu(\sigma)]= \int_{\WW_{p+1}} 
d^{p+1}\sigma\;\sqrt{-\gamma}\, |\mathbf{k}|^n\,
\eeq
by considering variations of $X^\mu$ in directions transverse
to the worldvolume.

We can write it in a form that is
particularly practical for obtaining the blackfold equations in specific
calculations. First observe that the asymptotic time, conjugate to the
vector $\xi$, is related to proper time $t$ on the worldvolume by a factor of
$R_0$. If
we take the interval for the (trivial) integration over asymptotic time
to have finite length $\beta$, then
\beq
\int_{\WW_{p+1}} d^{p+1}\sigma\;\sqrt{-\gamma}\, |\mathbf{k}|^n
=\beta\int_{\mathcal{B}_p} dV_{(p)}\;R_0\, |\mathbf{k}|^n\,.
\eeq
Using now \eqref{modk} we find
\beqa\label{statact2}
I[X^\mu(\sigma)]&=&\beta\int_{\mathcal{B}_p} dV_{(p)}\; 
R_0^{n+1} (1-V^2)^{\frac{n}{2}}\nonumber\\
&=&\beta\int_{\mathcal{B}_p} dV_{(p)}\;R_0(\sigma)\,
\left(R_0^2(\sigma)-\sum_i \Omega_i^2
R_i^2(\sigma)\right)^{\frac{n}{2}}\,.
\eeqa
We emphasize again that 
$R_a$ are among the worldvolume field variables $X^\mu(\sigma)$. Of
course these enter as well through $dV_{(p)}$.\footnote{In \cite{Emparan:2009cs}
it was asserted that the
blackfold equations can be derived from the action $\int_{\WW_{p+1}}
d^{p+1}\sigma\sqrt{-\gamma}T^{ab}\gamma_{ab}$. This is not true in
general, but is correct in the stationary case since
$T^{ab}\gamma_{ab}\propto r_0^n\propto |\mathbf{k}|^n$.}

\subsection{First law}

Using eqs.~\eqref{AH}, \eqref{mass}, \eqref{angmom}, it is
straightforward to check that
the action \eqref{statact2} is
\beq\label{thermact}
I=\beta\left(M-\sum_i \Omega_i J_i -\frac{\kappa}{8\pi G}A_H\right)\,.
\eeq
This identity holds for any embedding, not necessarily a solution to the
extrinsic equations.
Thus, if we
regard $M$, $J_i$ and $A_H$ as functionals of
$X^\mu(\sigma)$,
and consider variations at
fixed surface gravity and angular velocities, we have
\beq
\frac{\delta I}{\delta X^\mu}=0\quad 
\Leftrightarrow\quad
\frac{\delta M}{\delta
X^\mu}-\sum_i \Omega_i \frac{\delta J_i}{\delta X^\mu}-\frac{\kappa}{8\pi
G}\frac{\delta A_H}{\delta X^\mu}
=0\,.
\eeq
Hence, solutions of the blackfold equations satisfy the `equilibrium
state' version of the first law of black hole mechanics. Conversely, the
blackfold equations for stationary configurations can be obtained as the
requirement that the first law be satisfied. If we regard $\kappa$ and
$\Omega_i$ as Lagrange multipliers we may also say that blackfolds
extremize the horizon area for given mass and angular momenta.

In the Euclidean
quantum gravity approach to black hole thermodynamics it is natural to
take $\beta$ to be the
period of Euclidean time, $\beta=1/T$. Using
$\kappa A_H/8\pi G=TS$ and
eq.~\eqref{thermact} we see that $I$
is equal, up to a factor, to the
Gibbs free energy $G$,
\beq
\beta^{-1}I=G=M-\sum_i \Omega_i J_i -TS\,.
\eeq
Therefore \eqref{statact} can be identified as the effective action that
approximates, in the blackfold regime $r_0/R\ll 1$, the gravitational
Euclidean action of the black hole \cite{Gibbons:1976ue}.

\bigskip

It is also possible to find action functionals for general,
possibly time-dependent blackfolds by adapting the action principles
developed for perfect fluids \cite{Brown:1992kc}. However, the usefulness of these
actions, which must deal with constraints such as $u^2=-1$,
appears to be somewhat limited so we do not dwell on them.

\section{Gregory-Laflamme and correlated thermodynamic
instability in blackfolds}
\label{sec:GLinBF}

The blackfold approach must capture the perturbative dynamics of a black
hole when the perturbation wavelength $\lambda$ is long,
\beq
\lambda\gg r_0 \,.
\eeq
These perturbations can be either intrinsic variations in the
thickness $r_0$ and local velocity $u$, or extrinsic variations in the
worldvolume embedding geometry $X$. In general, these two kinds of
perturbations are coupled. A detailed analysis of the perturbations of
solutions to the blackfold equations and their stability will be
presented elsewhere. Here we extract some simple but
important
consequences for perturbations with wavelength
\beq\label{rangelambda}
r_0\ll \lambda \ll R\,,
\eeq
\ie perturbations for which the worldvolume looks essentially flat,
${K_{\mu\nu}}^\rho\approx 0$. In this case it is easy to see that the
intrinsic and extrinsic perturbations decouple. 

It is instructive to perform the analysis for a general perfect fluid
\eqref{perfluid}, and then particularize to the neutral blackfold
fluid
\eqref{blackepsP}. For simplicity we consider a fluid initially at rest
$u^a=(1,0\dots)$, with uniform equilibrium energy density $\vep$ and pressure $P$.
The flat worldvolume metric is parametrized, in `static gauge', using
orthonormal coordinates $X^0=t$, $X^i=z^i$, $i=1,\dots p$ and the
transverse coordinates $X^m$ are held at constant values.
Introduce small perturbations 
\beq
\delta\vep\,,\qquad
\delta P=\frac{dP}{d\vep}\delta\vep\,,\qquad
 \delta u^a=(0,v^i)\,,\qquad
\delta X^m=\xi^m\,,
\eeq
and work to linearized order in them.
The perturbed stress tensor is
\beq\label{deltT}
T^{tt}=\vep +\delta\vep\,,\qquad
T^{ti}=(\vep+P)v^i\,,\qquad
T^{ii}=P+\frac{dP}{d\vep}\delta\vep\,,
\eeq
and the extrinsic curvature
\beq
\delta {K_{ab}}^m=\partial_a \partial_b \xi^m\,.
\eeq

The extrinsic equations \eqref{extreqs} then become

\beq
\left(\vep \partial_t^2+P\partial_i^2 \right)\xi^m=0\,.
\eeq
Thus transverse, elastic oscillations of the brane propagate with speed
\beq\label{cT}
c_T^2=-\frac{P}{\vep}\,.
\eeq

The intrinsic equations \eqref{intreqs} are
\beq
\partial_t T^{tt}+\partial_i T^{it}=0\,,\qquad
\partial_t T^{ti} +\partial_j T^{ji}=0\,,
\eeq
which can be combined into
\beq
\partial_t^2 T^{tt}-\partial_{ij}T^{ij}=0\,.
\eeq
For \eqref{deltT} we find
\beq
\left(\partial_t^2-\frac{dP}{d\vep}\partial_i^2 \right)\delta\vep=0\,,
\eeq
so longitudinal, sound-mode oscillations of the fluid propagate with speed
\beq\label{cL}
c_L^2=\frac{dP}{d\vep}\,.
\eeq

These derivations of eqs.~\eqref{cT} and \eqref{cL} are hardly new: they
are conventional ways to obtain the speeds of elastic and sound waves ---
in fact they have been obtained in \cite{Carter:1989xk,Carter:1993wy}
for brane
dynamics. They have a remarkable consequence: a brane with a worldvolume
fluid equation
of state such that
\beq
\frac{P}{\vep}\frac{dP}{d\vep}>0
\eeq
has 
\beq
c_L^2\, c_T^2<0 
\eeq
and so is unstable to either longitudinal or transverse oscillations
with wavelengths in the range \eqref{rangelambda}. For instance this
happens in the simple case $P=w \vep$ with constant $w$, where the
interpretation is easy (we assume $\vep >0$): positive {\it tension} is
required for elastic stability, but positive {\it pressure} is needed to
prevent that the fluid clumps under any density perturbation.

Neutral blackfolds have
\beq\label{cbfold}
c_L^2=-c_T^2=-\frac{1}{n+1}\,.
\eeq
and therefore are generically unstable to longitudinal sound-mode
oscillations and stable to elastic
oscillations in the range of wavelengths \eqref{rangelambda}.

This instability is not unexpected. Black branes suffer from
the Gregory-Laflamme instability
\cite{Gregory:1993vy}, which makes the horizon radius vary as
\beq\label{delr0}
\delta r_0 \sim e^{\Omega t +ik_i z^i}\,. 
\eeq
Here $\Omega$ is positive real and thus the frequency is imaginary. The
threshold mode for the instability, with $\Omega=0$ and $k=\sqrt{k^i
k^i}\neq 0$, has `small' wavelength $\lambda=2\pi/k \sim r_0$ and
therefore
cannot be seen in the blackfold approximation. But the GL instability
extends to arbitrarily small $k$, \ie arbitrarily long wavelengths, and
when $k$ is very small it should be captured by the blackfold dynamics.

The sound-mode instability corresponds precisely to this long-wavelength
part, $\Omega,k\to 0$, of the GL instability. Observe that sound waves
in a blackfold produce $\delta\vep\sim \delta P \sim \delta r_0$
\ie variations in the horizon thickness. Eq.~\eqref{cbfold} tells us
that these are unstable, of the form \eqref{delr0} with dispersion
relation 
\beq\label{longGL}
\Omega=\frac{1}{\sqrt{n+1}}\;k\,.
\eeq
A simple inspection of the slope at the origin in figure~1 of
\cite{Gregory:1993vy} shows good numerical agreement with
\eqref{longGL}. We leave a more precise derivation of this equation from
a detailed GL-type analysis to future work. Note also that the
dispersion relation \eqref{longGL} indicates that the collective
coordinate $r_0$ is a {\it ghost} (\ie its effective Lagrangian
$-c_L^{-2}(\partial_t\ln r_0)^2 + (\partial_{z_i}\ln r_0)^2$ has the
`wrong sign' for the
kinetic term).\footnote{The GL mode at threshold is instead tachyonic,
since its dispersion relation has imaginary mass.}

Moreover, observe that the Gibbs-Duhem relation
$dP=sd\TT$, from \eqref{firstlaw} and \eqref{duhem}, implies in general that
\beq
\frac{dP}{d\vep}=s\frac{d\TT}{d\vep}=\frac{s}{c_v}
\eeq
where $c_v$ is the isovolumetric specific heat. Thus the black brane is
dynamically unstable (to long-wavelength
GL modes) if and only if it is locally
thermodynamically unstable, $c_v<0$. This is precisely the content of the
`correlated stability conjecture' of Gubser and Mitra
\cite{Gubser:2000mm}. In fact our
method gives a quantitative expression for the
unstable dynamical frequency in terms of local thermodynamics as
\beq
\Omega=\sqrt{\frac{s}{|c_v|}}\;k\,,
\eeq
which as far as we know is a new result.

\medskip

The ordinary derivation of the GL instability involves a complicated
analysis of linearized gravitational perturbations of a black brane and
the numerical resolution of a boundary value problem for a differential
equation (which is moreover compounded at small $k$ since larger grids
are required to avoid finite-size problems). Here we have shown that the
long-wavelength component of the instability, \eqref{longGL}, can be
obtained by an almost trivial calculation of the sound speed in a
fluid\footnote{Observe that this is {\it not} a Jeans instability of the
fluid (which has sometimes been suggested as possibly related to the GL
instability) since gravitational forces within the fluid are entirely
absent in our analysis.}. In addition, the correlation between dynamical
and thermodynamical stability follows as an elementary consequence of
the thermodynamics of the effective fluid. In our opinion these results
are striking evidence of the power of the blackfold approach.

\section{Closing remarks}
\label{sec:discussion}

The formalism we have developed bears relation to two different earlier
effective descriptions of black hole dynamics. The extrinsic part is a
generalization to $p$-branes of the effective worldline formalism for
small black holes, whose size $r_0$ is much smaller than the length
scale $R=1/(\mathrm{acceleration})$ of their trajectories or the
wavelength of the gravitational radiation they emit
\cite{Poisson:2003nc,Gralla:2008fg,cleft}. The intrinsic part is similar to other
fluid-dynamical formalisms for horizon fluctuations, such as the
membrane paradigm and the fluid/AdS-gravity correspondence.

With respect to the first one, note that our formalism allows to
consider time-dependent situations, which typically involve the emission
of gravitational waves. This can be obtained by coupling the blackfold
effective stress tensor to the quadrupole formula for gravitational
radiation. This is also common in studies of gravitational wave emission
from small black holes and from cosmic strings. However, accounting for
the backreaction of this radiation on the blackfold requires going
beyond the generalized-geodesic approximation and dealing with the
notoriously subtle problem of the gravitational self-force
\cite{backreact}.

To relate the fluid/AdS-gravity correspondence to our approach take,
instead of a neutral black brane, a near-extremal D3-brane. The
blackfold formalism can be applied to it, too\footnote{Charged
blackfolds will be discussed elsewhere.}. The small scale corresponds to
the charge-radius $r_q$ of the D3-brane (which is much larger than the
non-extremality length $r_0$), and in the overlap-zone $r_q \ll r \ll R$
where the blackfold effective stress tensor is computed, the metric is
flat up to small corrections in $r_q/R$. The blackfold method here could
be regarded as an extension of the DBI approach to describe thermally
excited worldvolumes. The difference with respect to neutral branes is
that there is a region near the horizon where excitations with long
wavelengths $\gg r_q$ are localized. One can take a limit, Maldacena's
decoupling limit, to decouple all the far-zone effects from them. This
region is asymptotic to AdS$_5\times S^5$ with radius $r_q$, and
far-zone effects are absent since they would give rise to
non-normalizable modes in AdS and change the boundary geometry.
Integrating the degrees of freedom in the asymptotically AdS region one
gets only intrinsic, purely hydrodynamical collective modes. The
effective stress tensor thus obtained is again of quasilocal type and in
fact is the holographic stress tensor in AdS
\cite{Balasubramanian:1999re}. This is in principle different than the
one in the blackfold approach (which is computed in an asymptotically
flat overlap-zone) but clearly they are related. For instance, at the
perfect fluid level they are the same. The main point is that in the
fluid/AdS-gravity correspondence there is no extrinsic worldvolume
dynamics, nor far-zone backreaction. But the less general dynamics comes
with a number of advantages: first, it makes much simpler to compute
higher-derivative corrections to the perfect-fluid dynamics. Second, the
charge near extremality eliminates the sound-mode instability. Finally,
this specific correspondence can be argued to describe the
hydrodynamical regime of a strongly-coupled quantum Yang-Mills theory.
For a generic blackfold it is not known whether there is a useful dual
interpretation of that sort.

Connections to the membrane paradigm are also suggestive but remain
somewhat less precise. The membrane paradigm can be formulated for generic
black holes, including vacuum ones, in terms of an effective fluid with
a stress tensor of quasilocal type \cite{moremem}. But in the membrane
paradigm the boundary where the fluid lives is taken to lie right above
the horizon, and not in an intermediate-asymptotic region (overlap zone)
as in the blackfold approach. Perhaps the membrane-paradigm stress
tensor and the one for blackfolds can be related through Komar
integrals, at least in vacuum gravity. At any rate the membrane paradigm
seems to capture only what we refer to as intrinsic fluid dynamics, and
it is not clear to us how it could deal with the extrinsic embedding
dynamics of a black brane.


\section*{Acknowledgments} We thank Joan Camps for many useful
discussions during collaboration on another strand of this program. TH
also thanks Shiraz Minwalla for useful discussions. 
RE, TH and NO are grateful to the Benasque Center for Science for
hospitality and a stimulating environment during the Gravity Workshop
in July 2009, and they thank the participants there for very useful feedback 
and discussions. RE was supported by DURSI 2005 SGR
00082 and 2009 SGR 168, MEC FPA 2007-66665-C02 and CPAN CSD2007-00042
Consolider-Ingenio 2010. 
TH was supported by the Carlsberg foundation. 
VN was supported by an Individual Marie Curie Intra-European Fellowship and by
ANR-05-BLAN-0079-02 and MRTN-CT-2004-503369, and CNRS PICS {\#} 3059,
3747 and 4172. 

\begin{appendix}

\section{Geometry of embedded submanifolds}
\label{app:extrcurv}

We collect here several relevant definitions and results on the geometry of
submanifold embeddings. Some aspects are more extensively discussed in
\cite{Carter:1992vb}.

\subsection{Extrinsic curvature}

Assume the submanifold $\WW$ is embedded as $X^\mu(\sigma^a)$.
The pull-back of the spacetime metric onto $\WW$ is
\beq\label{gammaalbe2}
\gamma_{ab}=g_{\mu\nu}\partial_a X^\mu \partial_b
X^\nu\,.
\eeq
The first fundamental tensor of the surface is then
\beq
h^{\mu\nu}=\gamma^{ab}\partial_a X^\mu\partial_b X^\nu\,.
\eeq
It follows easily
 that 
\beq\label{projdx}
{h^\mu}_\nu\partial_a X^\nu=\partial_a X^\mu\,,
\eeq
and
\beq\label{projh}
h^\mu{}_\nu h^\nu{}_\rho =h^\mu{}_\rho
\eeq
so $h^\mu{}_\nu$ projects tensors onto directions tangent to $\WW$.
Decomposing the metric as
\beq
g_{\mu\nu}=h_{\mu\nu}+\perp_{\mu\nu}\,,
\eeq
we obtain the orthogonal projection tensor $\perp_{\mu\nu}$,
\beq\label{projdx2}
\perp_{\mu\nu}\partial_a X^\mu=0\,,\qquad 
{\perp_\mu}^\nu {\perp_\nu}^\rho ={\perp_\mu}^\rho
\,.
\eeq

The extrinsic curvature tensor can be defined as
\begin{eqnarray}
\label{defK}
{K_{\mu\nu}}^\rho = {h^\lambda}_\mu
{h^\sigma}_\nu \nabla_\lambda {h^\rho}_\sigma=-{h^\lambda}_\mu
{h^\sigma}_\nu \nabla_\lambda {\perp^\rho}_\sigma\,.
\end{eqnarray}
The tangentiality of the first two indices and orthogonality of the last,
\beq
{\perp_{\mu}}^\nu {K_{\sigma\nu}}^\rho=
{\perp_{\sigma}}^\nu {K_{\nu\mu}}^\rho=
{h_{\nu}}^\rho {K_{\sigma\mu}}^\nu=0\,
\eeq
follows easily from this definition and the projector property \eqref{projh}.

Following \cite{Carter:2000wv}, it is convenient to introduce the tangential
covariant derivative
\beq
\overline\nabla_\mu ={h_\mu}^\nu\nabla_\nu\,,
\eeq
so 
\beq
{K_{\mu\nu}}^\rho={h_\nu}^\sigma \overline\nabla_\mu {h_\sigma}^\rho\,.
\eeq
Applying the tangential derivative on \eqref{projh} one obtains
\beq\label{symmK}
2K_{\mu(\nu\rho)}=\overline\nabla_\mu h_{\nu\rho}=-\overline\nabla_\mu
\perp_{\nu\rho}\,.
\eeq

Let $v$ be any vector tangent to $\WW$. Then \cite{Carter:1992vb}
\beqa\label{vvK}
v^\mu v^\nu {K_{\mu\nu}}^\rho&=&-v^\mu v^\nu \nabla_\nu{\perp_\mu}^\rho
=-v^\nu\nabla_\nu(v^\mu{\perp_\mu}^\rho)+{\perp^\rho}_\mu v^\nu\nabla_\nu v^\mu
\nonumber\\
&&={\perp^\rho}_\mu \dot v^\mu
\eeqa
where
\beq
\dot v^\mu=v^\nu\nabla_\nu v^\mu\,.
\eeq
Now let $N$ be any vector orthogonal to $\WW$. Then
\beq\label{NK}
N_\rho {K_{\mu\nu}}^\rho=N_\rho {h_\nu}^\sigma \oln_\mu {h_\sigma}^\rho=
-{h_\nu}^\rho \oln_\mu N_\rho\,.
\eeq

The symmetry 
\beq\label{Ksymm}
{K_{[\mu\nu]}}^\rho=0
\eeq 
follows from the integrability of the subspaces orthogonal to
$\perp_{\mu\nu}$. To prove this, assume that the latter is true, namely
that there is a submanifold $\WW$ defined by a set of equations
$f_{(i)}(X)=0$ such that
$df_{(i)}$ are
a basis of one-forms normal to the submanifold. Any one-form normal
to $\WW$ is a linear combination of them so the subspace
orthogonal to it is also integrable. It is always possible to choose a
one-form $N$ orthogonal to this subspace such that
\beq
N_\mu=\partial_\mu f(X)=\nabla_\mu f(X)\,,
\eeq
so, using \eqref{NK},
\beq
N_\rho {K_{\mu\nu}}^\rho=
-{h_{\nu}}^\sigma {h_{\mu}}^\rho \nabla_\sigma \nabla_\rho f\,,
\eeq
which is manifestly symmetric under $\mu\leftrightarrow \nu$. The
converse statement that \eqref{Ksymm} implies the integrability
of the orthogonal subspace, can also be proven by a straightforward
application of Frobenius's theorem \cite{Carter:1992vb}.

Background tensors ${t_{\mu_1 \mu_2\dots}}^{\nu_1\nu_2\dots}$ can be
pulled-back onto worldvolume tensors ${t_{a_1
a_2\dots}}^{b_1 b_2\dots}$ using
$\partial_a X^\mu$ as
\beq\label{wvtensors}
{t_{a_1 a_2\dots}}^{b_1 b_2\dots}=
\partial_{a_1} X^{\mu_1}\partial_{a_2} X^{\mu_2} \cdots
\partial^{b_1} X_{\nu_1}\partial^{b_2} X_{\nu_2}\cdots
{t_{\mu_1 \mu_2\dots}}^{\nu_1\nu_2\dots}\,,
\eeq
where
\beq
\partial^{b} X_{\nu}=\gamma^{bc}h_{\nu\rho}\partial_{c} X^{\rho}\,.
\eeq

Observe that even when ${t_{\mu_1 \mu_2\dots}}^{\nu_1\nu_2\dots}$ is a
background tensor with indices parallel to $\WW$, in general $\oln_\mu
{t_{\mu_1 \mu_2\dots}}^{\nu_1\nu_2\dots}$ has both parallel and
orthogonal components. The parallel projection along all indices is
related to the worldvolume covariant derivative $D_a{t_{a_1
a_2\dots}}^{b_1 b_2\dots}$ as in \eqref{wvtensors}. This can be shown by
using the equation that relates the connection coefficients for each
metric, $\Gamma^\rho_{\mu\nu}$ and $\left\{ {c\atop a\,b}\right\}$,
\beq\label{christoffels}
\partial_a X^\mu\partial_b X^\nu {h^\sigma}_\rho \Gamma^\rho_{\mu\nu}
=\partial_c X^\sigma \left\{ {\textstyle{c\atop a\,b}}\right\}-
{h^\sigma}_\rho \partial_a\partial_b X^\rho\,,
\eeq
which can be proven by direct substitution of the definitions of each
term involved. 

In
particular, the divergences of tensors are related as
\beq\label{divs}
{h^{\nu_1}}_{\mu_1}\cdots \oln_\rho t^{\rho{\mu_1}\dots}
=\partial_{a_1}X^{\nu_1}\cdots D_c t^{c a_1\dots}\,.
\eeq
Such relations allow to dispense with the use of worldvolume coordinate
tensors and derivatives in most formal manipulations. However,
worldvolume
coordinates are very practical for explicit calculations and also allow
us to highlight the distinction between intrinsic (parallel to
 $\WW$) and extrinsic (orthogonal to $\WW$) equations.

Let us now consider the divergence of a totally antisymmetric
tensor $J$ (such as a current associated to a gauge form field) parallel to
the worldvolume. It is easy to
show that
\beq
{\perp^\rho}_{\mu_1}\oln_\mu J^{\mu\mu_1\dots}=0\,
\eeq
holds as an identity.
This implies that the conservation equation
\beq\label{nablaJ}
\oln_\mu J^{\mu\mu_1\dots}=0
\eeq
is equivalent to the worldvolume conservation equation
\beq
D_a J^{a a_1\dots}=0\,,
\eeq
\ie the orthogonal component of the current conservation
equation \eqref{nablaJ} does not yield
any additional equation. 
In particular, for a 1-form current one has
\beq
\oln_\mu J^\mu =D_a J^a\,,
\eeq
and continuity of charge is only meaningful as an intrinsic equation.
This is in contrast to the conservation of the
stress energy tensor, where the orthogonal component gives 
independent extrinsic equations \eqref{extreqs}.

Let us now obtain more explicit expressions for the pull-back of the
extrinsic curvature tensor onto
$\WW$ in terms of $X^\mu(\sigma)$,
\begin{equation}
{K_{ab}}^\rho = \partial_a X^\mu \partial_b X^\nu
{K_{\mu\nu}}^\rho=
-\partial_a X^\mu \partial_b X^\nu \nabla_\mu {\perp_{\nu}}^\rho\,.
\eeq
The property \eqref{projdx2} implies
\beq
0=\partial_b X^\nu \nabla_\nu\left(
{\perp_{\sigma}}^\rho\partial_a X^\sigma\right)=
-{K_{ab}}^\rho
+{\perp_{\sigma}}^\rho\partial_b X^\nu 
\nabla_\nu\left(\partial_a X^\sigma\right)\,.
\eeq
Expanding the covariant derivative in the last term and using 
$\partial_b X^\nu \partial_\nu=\partial_b$ we find
\beq\label{niceextr1}
{K_{ab}}^\rho=
{\perp_\sigma}^\rho\left(\partial_a\partial_b X^\sigma
+\Gamma^{\sigma}_{\mu\nu}\partial_a X^\mu\partial_b X^\nu\right)\,,
\eeq
which is reminiscent of the expression for the acceleration (deviation
from self-parallel transport) of a curve --- indeed \eqref{vvK} makes
this even more explicit.
An alternative expression with this same
feature can be obtained
by performing some manipulations:
\beqa
{\perp_\sigma}^\rho\partial_a\partial_b X^\sigma&=&
\partial_a\partial_b X^\rho-{h^\rho}_\sigma \partial_a\partial_b X^\sigma
\nonumber\\
&=&\partial_a\partial_b X^\rho  
-\left\{ {\textstyle{c\atop a\,b}}\right\}\partial_c X^\rho
+\partial_a X^\mu \partial_b X^\nu {h^\rho}_\sigma
\Gamma^\sigma_{\mu\nu}\nonumber\\
&=&D_a \partial_b X^\rho + \partial_a X^\mu \partial_b X^\nu {h^\rho}_\sigma
\Gamma^\sigma_{\mu\nu}\,.
\eeqa
In the second line we have used \eqref{christoffels}. Inserting
the last expression into \eqref{niceextr1} we find
\begin{equation}
\label{niceextr} {K_{ab}}^\rho = D_a
\partial_b X^\rho + \Gamma^\rho_{\mu \nu} \partial_a X^\mu
\partial_b X^\nu\,.
\end{equation}

\subsection{Variational calculus}

Consider a congruence of curves
with tangent vector $N$, that
intersect
$\WW$ orthogonally
\beq
N^\mu h_{\mu\nu}=0\,,\qquad N^\mu \perp_{\mu\nu}=N_\nu\,,
\eeq 
and Lie-drag $\WW$ along these curves. 
The congruence is arbitrary, other than
requiring it to be smooth in a finite
neighbourhood of $\WW$, so this realizes arbitrary smooth deformations
of the worldvolume $X^\mu \to X^\mu +N^\mu$.

Consider now the Lie derivative of $h_{\mu\nu}$ along $N$. In general,
\beq
\Ld_N h_{\mu\nu}=N^\rho \nabla_\rho h_{\mu\nu}+h_{\rho\nu}\nabla_\mu N^\rho
+h_{\mu\rho}\nabla_\nu N^\rho\,.
\eeq
Then
\beq
{h_\mu}^\lambda {h_\nu}^\sigma \Ld_N h_{\lambda\sigma}=
{h_\mu}^\lambda {h_\nu}^\sigma N^\rho \nabla_\rho h_{\lambda\sigma}+
h_{\rho\nu}\oln_\mu N^\rho +h_{\mu\rho}\oln_\nu N^\rho\,.
\eeq
The first term in the rhs is zero:
\beqa
{h_\mu}^\lambda {h_\nu}^\sigma N^\rho \nabla_\rho h_{\lambda\sigma}&=&
-{h_\mu}^\lambda {h_\nu}^\sigma N^\rho \nabla_\rho \perp_{\lambda\sigma}\nonumber\\
&=&{h_\mu}^\lambda \perp_{\lambda\sigma} N^\rho \nabla_\rho {h_\nu}^\sigma=0\,,
\eeqa
and the other two terms can be rewritten using \eqref{NK}, so
\beq
N_\rho {K_{\mu\nu}}^\rho=
-\frac{1}{2}{h_\mu}^\lambda {h_\nu}^\sigma \Ld_N h_{\lambda\sigma}\,.
\eeq
This implies
\beq
N_\rho K^\rho=-\frac{1}{2}h^{\mu\nu} \Ld_N h_{\mu\nu}=
-\frac{1}{\sqrt{|h|}}\Ld_N \sqrt{|h|}\,,
\eeq
where $h=\det h_{\mu\nu}$. These equations generalize well-known expressions for the extrinsic curvature of
a codimension-1 surface. The last one allows to derive the equation for
a minimal-volume submanifold:
\beq
\mathit{Vol}=\int_{\WW} \sqrt{|h|} \quad \Rightarrow \quad \delta_N \mathit{Vol}
=-\sqrt{|h|}\; N_\rho K^\rho
\eeq
\ie for variations along an arbitrary orthogonal direction $N$, minimal
(actually, extremal) volume requires $K^\rho=0$. This is of course the
variational principle
for Nambu-Goto-Dirac branes.

Consider now a more general functional
\beq
I=\int_{\WW} \sqrt{|h|}\, \Phi
\eeq
where $\Phi$ is a worldvolume function. Then
\beq
\delta_N I=\Ld_N\left(\sqrt{|h|}\,\Phi\right)=
\sqrt{|h|}\left(-N_\rho K^\rho \Phi+N^\rho\partial_\rho\Phi\right)\,.
\eeq
Since $N$ is an arbitrary orthogonal vector we have
\beq\label{varI}
\delta_N I=0 \quad 
\Leftrightarrow \quad K^\rho =\perp^{\rho\mu}\partial_\mu \ln \Phi\,.
\eeq

\end{appendix}

\end{document}